\documentclass[11pt]{article}

\usepackage{fullpage,amssymb,graphicx,amsfonts,amsmath,mathrsfs}
\usepackage{setspace}
\usepackage{natbib}
\usepackage[]{color}

\begin{document}

% Title of paper
\title{Bayesian nonparametric mean residual life
regression}

% List of authors, with corresponding author marked by asterisk
\author{VALERIE POYNOR \\
\textit{Department of Mathematics, California State University, Fullerton} \\[2pt]
ATHANASIOS KOTTAS$^\ast$\\
\textit{Department of Statistics,
University of California, Santa Cruz}
\\[1pt]
% E-mail address for correspondence
{thanos@soe.ucsc.edu}}

% Running headers of paper:
\markboth%
% First field is the short list of authors
{V. Poynor and A. Kottas}
% Second field is the short title of the paper
{Bayesian nonparametric mean residual life
regression}
\date{}
\vspace{-2pt}

\maketitle

\begin{center} 
{SUMMARY}
\end{center}

{The mean residual life function is a key functional for a survival distribution. It has a practically useful interpretation as the expected remaining lifetime given survival up to a particular time point, and it also characterizes the survival distribution. However, it has received limited attention in terms of inference methods under a probabilistic modeling framework. We seek to provide general inference methodology for mean residual life regression. We employ Dirichlet process mixture modeling for the joint stochastic mechanism of the covariates and the survival response. This density regression approach implies a flexible model structure for the mean residual life of the conditional response distribution, allowing general shapes for mean residual life as a function of covariates given a specific time point, as well as a function of time given particular values of the covariates. We further extend the mixture model to incorporate dependence across experimental groups. This extension is built from a dependent Dirichlet process prior for the group-specific mixing distributions, with common atoms and weights that vary across groups through latent bivariate Beta distributed random variables. We discuss properties of the regression models, and develop methods for posterior inference. The different components of the methodology are illustrated with simulated data examples, and the model is also applied to a data set comprising right censored survival times.
}

\newpage
 
\section{Introduction}
\label{s:intro}

The mean residual life (MRL) function of a continuous positive-valued random variable, $T$, provides 
the expected remaining lifetime given survival up to time $t$, $m( t) = \text{E}(T - t \mid T>t)$. 
Its definition requires that $T$ has finite mean, which is given by $\text{E}(T) = m(0)$. 
The MRL function can be defined through the survival function, $S(t)=\text{P}(T>t)$, in particular, 
$m(t) =\left[\int_t^\infty S(u) \text{d}u\right]/S(t)$, with $m(t) \equiv 0$ when $S(t) = 0$. 
Conversely, the survival function is defined through the MRL function via 
$S(t)=\{ m(0)/m(t) \} \exp[ -\int_{0}^{t} \{ 1/m(u) \} \text{d}u]$ \citep{hall-wellner:1981}, 
and thus the MRL function characterizes the survival distribution. Given this property 
and its useful interpretation, the MRL function is of practical importance in a variety 
of fields, such as reliability, medicine, and actuarial science.

Often associated with the survival times is a set of covariates, ${\boldsymbol x}$. 
The MRL regression function at a specified set of covariate values is given by:
\begin{eqnarray} 
m( t \mid {\boldsymbol x}) = \text{E}(T - t \mid T>t, {\boldsymbol x}) = 
\frac{\int_t^\infty S(u \mid {\boldsymbol x}) \, \text{d}u}{S(t \mid {\boldsymbol x})}  \label{eqn:mrl}
\end{eqnarray}
provided $\text{E}(T \mid {\boldsymbol x})<\infty$. In the regression setting, it is of interest 
to develop modeling that allows flexible MRL function shapes over the covariate space. Note that MRL 
regression involves study of covariate effects on a function over time. It is of interest to study 
how MRL regression relationships change across different time points (mean regression corresponds 
to $t=0$), as well as how the MRL function evolves across different covariate values.

Classical estimation methods for MRL regression have been primarily derived from the proportional 
MRL model, $m(t) =\gamma \, m_0(t)$, where $\gamma >0$ and $m_0(t)$ is a baseline MRL function 
\citep{oakes-dasu:1990}. 
%If $\gamma <1$, the survival function, $S(t)$, associated with $m(t)$ 
%is proper for any proper MRL function $m_0(t)$. Alternatively, $S(t)$ is proper for all 
%$\gamma > 0$ if and only if $m_0(t)$ is nondecreasing. 
\cite{maguluri-zhang:1994} extend the proportional MRL model 
to incorporate covariates, such that the MRL regression function is 
$m_0(t) \exp({\boldsymbol \beta}^T{\boldsymbol x})$, where ${\boldsymbol \beta}$ are the 
regression coefficients.
%and $m_0(t)$ is the unknown baseline MRL function.  
They propose two estimators for ${\boldsymbol \beta}$ under fully observed survival responses.  
%\cite{chen-cheng:2005} expand the estimation methods under the semiparametric proportional MRL model 
%to include censored responses, \cite{chen-wang:2016} address the case with the censoring indicators 
%missing at random, and \cite{bai-etal:2016} account for right censored length-biased data. 
\cite{chen-cheng:2005}, \cite{chen-wang:2016} and \cite{bai-etal:2016} expand the estimation 
methods under different forms of censoring. 
Alternative to the proportional MRL model structure, \cite{chen:2006} and \cite{chen-cheng:2006} 
develop additive MRL models with the MRL regression function given by 
$m_0(t) + {\boldsymbol \beta}^T{\boldsymbol x}$. \cite{sun-zhang:2009} generalize such models 
to $g(m_0(t) + {\boldsymbol \beta}^T{\boldsymbol x})$, where $g$ is a pre-specified link function.
This model is further extended in \cite{sun-etal:2012} to 
incorporate time-dependent regression parameters in a linear fashion. \cite{jin-etal:2020} 
and \cite{Ma_et_al_2020} consider generalized MRL models under different sampling scenarios.
%consider the generalized MRL model under the case-cohort and nested case-control design. 
%These methods are more general than the basic approach of linking covariates through parameters 
%of a parametric MRL function. However, they are still 
Classical semiparametric estimation methods are restricted by the: proportional or additive 
form of the MRL regression function; the parametric introduction of covariate effects; and the fact 
that they are not based on fully probabilistic model settings, thus requiring asymptotic 
arguments for uncertainty quantification.

Survival analysis is one of the first application areas for nonparametric Bayesian (NPB) modeling and 
inference methods. The relevant literature includes several prior probability models for cumulative 
hazard functions, hazard functions, survival functions, or survival densities; see, e.g., 
\cite{Ibrahim_2001}, \cite{Phadia2013}, and \cite{Mueller2015}. However, to our knowledge, there is 
no work in the NPB literature that explores modeling and inference for the MRL function in the 
presence of covariates.

Our objective is to develop an NPB modeling framework that lends full and flexible inference for MRL 
regression. %within and across the covariate space. 
To accommodate the regression setting, we extend our earlier work on inference for MRL functions, 
based on Dirichlet process (DP) mixture priors for the survival distribution \citep{poynor-kottas:2019}. 
To this end, we employ DP mixture modeling for the joint stochastic mechanism of the survival 
response and the covariates, from which general inference for MRL regression emerges through the 
implied conditional response distribution. This DP mixture density regression approach was proposed 
by \cite{muller-etal:1996} for real-valued responses, and it has been applied and elaborated under 
different settings; see, e.g., \cite{MDAKbook} for a review and relevant references.
%see, e.g., \cite{muller-quintana:2010}, \cite{taddy-kottas:2010}, \cite{wade-etal:2014}, 
%\cite{papageorgiou-etal:2015}, \cite{deyoreo-kottas:2018}, and (REF on density autoregression).
In our context, a key modeling choice involves the mixture kernel that corresponds to the survival 
responses, for which we argue for the gamma density. Even though we do not model directly the MRL 
function of the response distribution, the implied prior model for the MRL function given the 
covariates has an appealing structure as a locally weighted mixture of the kernel MRL functions, 
with weights that depend on both time and the covariates.

For problems with a small to moderate number of random covariates, the density regression modeling 
approach is attractive in terms of its inferential flexibility. At the same time, survival data 
typically comprise responses from subjects assigned to different experimental groups, e.g., control 
and treatment groups. 
%The treatment indicator can not be meaningfully incorporated into the joint response-covariate 
%mixture model as an additional component of the mixture kernel. 
Since the treatment indicator is not a random covariate, we extend the model to allow distinct 
mixing distributions for the different groups, which are dependent in their nonparametric prior.
We develop this extension in the context of two groups, using a dependent Dirichlet process 
prior \citep{maceachern:2000,DDPreview} for the group-specific mixing distributions.  
The general model retains the local weighted mixture structure, with local adjustment provided 
by weights that depend on time, covariate values, as well as the treatment groups.

The outline of the paper is as follows. Section~\ref{sec:curve_reg} develops the density 
regression approach to modeling and inference for MRL regression. In Section~\ref{sec:ddp}, we 
present the model elaboration to incorporate survival data from different experimental groups, 
including illustrations from two synthetic data examples. In Section~\ref{sec:lung_data}, we provide 
a detailed analysis of a standard data set from the literature on right censored survival times 
for patients with small cell lung cancer. Finally, Section~\ref{sec:conc} concludes with a summary.

\section{Mean residual life regression}
\label{sec:curve_reg}

\subsection{Model formulation}
\label{DPM_for_MRL}

For survival regression problems with a small/moderate number of random covariates, it is 
meaningful to model the joint distribution of the survival response and covariates. In our 
context, a key benefit of this modeling strategy involves the implied MRL function of the 
conditional response distribution. In particular, we obtain general MRL regression relationships 
across different time points, and flexible MRL functions across different covariate values.

Let ${\boldsymbol x}$ be a vector of random covariates and $T$ the positive-valued survival 
response variable. We model the joint response-covariate density using a DP mixture model:
\begin{eqnarray}\label{eqn:dpmm}
f(t, {\boldsymbol x} \mid G) \, = \, 
\int k(t, {\boldsymbol x} \mid {\boldsymbol \theta}) \, \text{d}G( {\boldsymbol \theta});  
\ \ \  G \sim \text{DP}(\alpha, G_0) 
\end{eqnarray}
where $k(t, {\boldsymbol x} \mid {\boldsymbol \theta})$ is the joint kernel density for survival 
time and covariates, and the mixing distribution, $G$, is assigned a DP prior \citep{ferguson:1973}. 
The model is completed with hyperpriors for the DP precision parameter, $\alpha$, and for (some of) 
the parameters of the baseline (centering) distribution $G_{0}$. Under the DP constructive 
definition \citep{sethuraman:1994}, a realization $G$ from DP$(\alpha, G_0)$ is almost surely 
of the form $\sum_{l=1}^\infty w_l \, \delta_{\boldsymbol{\theta}_l}$, where the atoms are 
independently and identically distributed (i.i.d.) from the baseline distribution,
$\boldsymbol{\theta}_l\stackrel{\text{i.i.d.}}{\sim} G_{0}$, with the weights constructed through 
stick-breaking: $w_1= v_1$, and $w_l=$ $v_l\prod_{r=1}^{l-1} (1-v_r)$, for $l \geq 2$, 
where $v_l \stackrel{\text{i.i.d.}}{\sim} \text{Beta}(1,\alpha)$ (independently of 
the $\boldsymbol{\theta}_l$).

Hence, the density in (\ref{eqn:dpmm}) can be re-written as $f(t, {\boldsymbol x} \mid G) =$
$\sum_{l = 1}^\infty w_l \, k(t, {\boldsymbol x} \mid {\boldsymbol \theta}_l)$. Directly from 
their definitions, the conditional response density can be expressed as $f(t \mid {\boldsymbol x}, G)=$
$\sum_{l=1}^{\infty}  q_l({\boldsymbol x}; {\boldsymbol \theta}_l) \, 
k(t \mid {\boldsymbol x}, {\boldsymbol \theta}_l)$, and the conditional survival function as 
\begin{equation}
\label{DPM_survival}
S(t \mid {\boldsymbol x}, G) \, = \, 
\sum_{l=1}^{\infty}  q_l({\boldsymbol x}; {\boldsymbol \theta}_l) \, 
S(t \mid {\boldsymbol x}, {\boldsymbol \theta}_l)
\end{equation}
where $q_l({\boldsymbol x}; {\boldsymbol \theta}_l) =$
$w_l \, k({\boldsymbol x} \mid {\boldsymbol \theta}_l) / \{\sum_{r=1}^\infty w_r \, k({\boldsymbol x} 
\mid {\boldsymbol \theta}_r)\} =$
$w_l \, k({\boldsymbol x} \mid {\boldsymbol \theta}_l) / f( {\boldsymbol x} \mid G)$.
Therefore, the conditional density and survival functions are represented as mixtures of 
the corresponding kernel functions with covariate-dependent mixture weights. 
Analogously, the mean regression function is given by $\text{E}(T \mid {\boldsymbol x}, G) =$ 
$\sum_{l=1}^\infty q_l({\boldsymbol x}; {\boldsymbol \theta}_l) \,
\text{E}(T \mid {\boldsymbol x},{\boldsymbol \theta}_l)$ (a sufficient condition for the conditional 
expectation to be finite is given in Section \ref{DPM_kernel}). The covariate-dependent mixture weights 
allow for local adjustment over the covariate space, thus enabling general shapes for the conditional 
response distribution and for the mean regression functional.

Importantly for our objective, this local mixture structure extends to the MRL functional. Using the 
form for the conditional survival function in (\ref{DPM_survival}) and the definition of the MRL 
regression function from (\ref{eqn:mrl}), we obtain
\begin{eqnarray}
\label{eqn:dpmm_mrl}
m(t \mid {\boldsymbol x}, G) \, = \, 
\frac{\int_t^\infty S(u \mid {\boldsymbol x},G) \, \text{d}u}{S(t \mid {\boldsymbol x},G)} \, = \, 
\sum_{l=1}^{\infty} q^{*}_{l}(t,{\boldsymbol x} ;{\boldsymbol \theta}_l) \, 
m(t \mid {\boldsymbol x},{\boldsymbol \theta}_l)
\end{eqnarray}
where 
\begin{equation}
\label{MRL_mixture_weights}
q^{*}_{l}(t, {\boldsymbol x} ; {\boldsymbol \theta}_l) \, = \,
\frac{ w_l \, k({\boldsymbol x} \mid {\boldsymbol \theta}_l) \, 
S(t \mid {\boldsymbol x}, {\boldsymbol \theta}_l) }
{ \sum_{r=1}^{\infty} w_{r} \, k({\boldsymbol x} \mid {\boldsymbol \theta}_r) \,
S(t \mid {\boldsymbol x}, {\boldsymbol \theta}_r) }
\end{equation}
and $m(t \mid {\boldsymbol x},{\boldsymbol \theta})$ is the MRL function of the mixture 
kernel conditional response distribution. (Implicit here is the assumption that, under 
the kernel distribution, $\text{E}(T \mid {\boldsymbol x},{\boldsymbol \theta}) < \infty$, for 
any ${\boldsymbol x}$). Therefore, our prior model for the MRL regression function admits a 
representation as a weighted sum of the conditional MRL functions associated with the kernel 
components, with weights that are dependent on both time and the covariate values. Important to 
note in the form of the mixture weights in (\ref{MRL_mixture_weights}) is that there are separate 
functions controlling the local adjustment over covariate values and time. Aside from the 
useful interpretation, expressions (\ref{eqn:dpmm_mrl}) and (\ref{MRL_mixture_weights}) suggest 
the model's capacity to capture non-standard MRL regression relationships over time, and 
general MRL function shapes across the covariate space.

\subsection{Mixture kernel specification}
\label{DPM_kernel}

A key aspect for the model formulation is the choice of the DP mixture kernel, 
$k(t, {\boldsymbol x} \mid {\boldsymbol \theta})$. A structured approach to specifying 
dependent kernel densities involves a marginal density for the covariates, 
$k({\boldsymbol x} \mid {\boldsymbol \theta}_{1})$, 
and a parametric regression model for $k(t \mid {\boldsymbol x},{\boldsymbol \theta}_{2})$, where 
${\boldsymbol \theta}=$ $({\boldsymbol \theta}_{1},{\boldsymbol \theta}_{2})$. For our data illustrations, 
we use the simpler form for the kernel density with independent components for the survival 
response and the covariates, $k(t, {\boldsymbol x} \mid {\boldsymbol \theta}) =$ 
$k({\boldsymbol x} \mid {\boldsymbol \theta}_{1}) \, k(t \mid {\boldsymbol \theta}_{2})$. 
In this case, the prior model in (\ref{eqn:dpmm_mrl}) becomes a weighted combination of the marginal 
kernel MRL functions, $m(t \mid {\boldsymbol \theta}_l)$, with weights that are still dependent 
on both time and covariate values through distinct functions, in particular, 
$S(t \mid {\boldsymbol \theta}_l)$ (replacing $S(t \mid {\boldsymbol x}, {\boldsymbol \theta}_l)$
in (\ref{MRL_mixture_weights})) and $k({\boldsymbol x} \mid {\boldsymbol \theta}_l)$, respectively. 
As can be seen from the model structure, 
%and also demonstrated with the data examples, 
such a kernel density form strikes a useful balance between model flexibility and complexity 
with respect to the dimensionality of the mixing parameter vector ${\boldsymbol \theta}_{2}$.

Regarding $k({\boldsymbol x} \mid {\boldsymbol \theta}_{1})$, when all the covariates are continuous, 
the multivariate normal density is a convenient choice, possibly after transformation for the values 
of some of the covariates. A normal kernel density can also accommodate ordinal categorical covariates 
through latent continuous variables \citep[e.g.,][]{MDAK2018}. Alternatively, categorical 
covariates (whether ordinal or nominal) can be incorporated by adding a corresponding component to the 
kernel in a product form, or if relevant, through marginal and conditional densities for the continuous 
and categorical covariates \citep[e.g.,][]{taddy-kottas:2010}.

A key consideration for the specification of $k(t \mid {\boldsymbol x},{\boldsymbol \theta}_{2})$
(or $k(t \mid {\boldsymbol \theta}_{2})$) is to ensure that the MRL function 
$m(t \mid {\boldsymbol x}, G)$ is well defined, that is, $\text{E}(T \mid {\boldsymbol x},G)$ must 
be (almost surely) finite, for any ${\boldsymbol x}$. The following lemma provides sufficient 
conditions in that direction.

\vspace{0.15cm}
\noindent 
{\bf Lemma.} Consider the DP mixture model in (\ref{eqn:dpmm}) with kernel 
$k(t, {\boldsymbol x} \mid {\boldsymbol \theta}_{1}, {\boldsymbol \theta}_{2}) =$
$k({\boldsymbol x} \mid {\boldsymbol \theta}_{1}) \, k(t \mid {\boldsymbol x},{\boldsymbol \theta}_{2})$, 
and with DP baseline distribution $G_{0}({\boldsymbol \theta}_{1},{\boldsymbol \theta}_{2})=$ 
$G_{10}({\boldsymbol \theta}_{1}) \, G_{20}({\boldsymbol \theta}_{2})$. For a generic set of covariate 
values ${\boldsymbol x}$, assume that: (a) $\text{E}(T \mid {\boldsymbol x},{\boldsymbol \theta}_{2}) =$
$\int_{\mathbb{R}^{+}} u \, k(u \mid {\boldsymbol x},{\boldsymbol \theta}_{2}) \, \text{d}u < \infty$;
and, (b) $\int \text{E}(T \mid {\boldsymbol x},{\boldsymbol \theta}_{2}) \, 
\text{d}G_{20}({\boldsymbol \theta}_{2}) < \infty$. Then, $\text{E}(T \mid {\boldsymbol x},G) < \infty$, 
almost surely. 

\vspace{0.1cm}
\noindent 
{\it Proof.}
Based on the DP constructive definition, 
$$
\text{E}(T \mid {\boldsymbol x},G) \, = \,
\sum_{l = 1}^{\infty} q_l({\boldsymbol x}; {\boldsymbol \theta}_{1l}) \,
\text{E}(T \mid {\boldsymbol x},{\boldsymbol \theta}_{2l}) \, = \,
\frac{\sum_{l = 1}^{\infty} w_{l} \, A_{{\boldsymbol x}}({\boldsymbol \theta}_{l})}
{f({\boldsymbol x} \mid G)}
$$
where $A_{{\boldsymbol x}}({\boldsymbol \theta}) =$
$\int_{\mathbb{R}^{+}} u \, k(u, {\boldsymbol x} \mid {\boldsymbol \theta}) \, \text{d}u =$
$k({\boldsymbol x} \mid {\boldsymbol \theta}_{1}) \,
\text{E}(T \mid {\boldsymbol x},{\boldsymbol \theta}_{2})$ $< \infty$, from assumption (a). 
Let $Z_{{\boldsymbol x}} =$ $\sum_{l = 1}^{\infty} w_{l} \, A_{{\boldsymbol x}}({\boldsymbol \theta}_{l})$. 
Using the monotone convergence theorem, and the independence between the DP atoms and weights, 
we have $\text{E}(Z_{{\boldsymbol x}}) =$ 
$\sum_{l = 1}^{\infty} \text{E}(w_{l}) \, \text{E}(A_{{\boldsymbol x}}({\boldsymbol \theta}_{l})) =$
$\text{E}(A_{{\boldsymbol x}}({\boldsymbol \theta}_{l}))$, since this expectation is free of $l$ as the 
${\boldsymbol \theta}_{l}$ are i.i.d. (from $G_{0}$). Moreover, 
$$
\text{E}(A_{{\boldsymbol x}}({\boldsymbol \theta}_{l})) = \int A_{{\boldsymbol x}}({\boldsymbol \theta}) \,
\text{d}G_{0}({\boldsymbol \theta}_{1},{\boldsymbol \theta}_{2}) =
\left\{ \int k({\boldsymbol x} \mid {\boldsymbol \theta}_{1}) \, \text{d} G_{10}({\boldsymbol \theta}_{1}) \right\}
\left\{ \int \text{E}(T \mid {\boldsymbol x},{\boldsymbol \theta}_{2}) \,
\text{d}G_{20}({\boldsymbol \theta}_{2}) \right\}
$$ 
which is finite from assumption (b). Since $Z_{{\boldsymbol x}}$ is a positive-valued random 
variable with finite expectation, $Z_{{\boldsymbol x}}$ is almost surely finite, and thus, 
$\text{E}(T \mid {\boldsymbol x},G) < \infty$, almost surely.

\vspace{0.15cm}

Note that defining $G_{0}$ through independent components for ${\boldsymbol \theta}_{1}$ and 
${\boldsymbol \theta}_{2}$ is a natural modeling strategy. Also, under independent kernel 
components for the response and covariates, 
$\text{E}(T \mid {\boldsymbol x},{\boldsymbol \theta}_{2})$ simplifies to the expectation of 
the kernel for survival time, and the second lemma condition becomes 
$\int \text{E}(T \mid {\boldsymbol \theta}_{2}) \, \text{d}G_{20}({\boldsymbol \theta}_{2}) < \infty$. 
For this model version, it is straightforward to verify the lemma conditions for the gamma density 
under the choice for $G_{20}$ discussed below. The gamma choice is unique in this respect among standard 
lifetime distributions in that it ensures existence of the mixture MRL function without the need for 
awkward restrictions on the parameter space for ${\boldsymbol \theta}_{2}$. Further support for the 
gamma kernel choice is provided by the fact that it generates both increasing and decreasing MRL 
functions (for shape parameter $< 1$ and $> 1$, respectively), its MRL function can be readily 
computed (see Section \ref{DPM_MRL_inference}), as well as by a denseness result for the MRL 
function of gamma mixture distributions, obtained under the setting without 
covariates \citep{poynor-kottas:2019}.

We use the following parameterization for the gamma density, 
$k(t \mid {\boldsymbol \theta}_{2}) \equiv$ 
$k(t \mid \eta,\phi) \propto$ $t^{e^\eta -1}\text{exp}(-e^\phi t)$, with 
$(\eta, \phi)\in \mathbb{R}^2$, to facilitate selection of a dependent $G_{20}(\eta,\phi)$ 
distribution, taken to be bivariate Gaussian. Finally, we note that the lemma conditions remain 
generally easy to verify if one wishes to extend the gamma kernel density to depend on covariates, 
for instance, such that its mean is extended to $\exp(\eta - {\boldsymbol x}^T{\boldsymbol \beta})$.

\subsection{Posterior inference}
\label{DPM_MRL_inference}

We obtain samples from the posterior distribution of the DP mixture model using the blocked Gibbs 
sampler \citep{ishwaran-james:2001}. In particular, the Markov chain Monte Carlo (MCMC) posterior 
simulation method builds from a truncation approximation to the mixing distribution, 
$G_{L}=$ $\sum_{l=1}^{L} p_{l} \, \delta_{\boldsymbol{\theta}_l}$, with 
$\boldsymbol{\theta}_l \stackrel{\text{i.i.d.}}{\sim} G_{0}$, for $l=1,...,L$, and $p_{l}=$
$w_{l}$, for $l=1,...,L-1$, with $p_L=$ $1- \sum_{l=1}^{L-1} p_{l}$. The truncation level $L$ 
can be chosen to any desired level of accuracy, using DP properties. For instance, the prior 
expectation for the partial sum of the DP weights, $\text{E}(\sum_{l=1}^{L} w_{l} \mid \alpha)=$ 
$1 - \{ \alpha/(\alpha + 1) \}^{L}$, can be averaged over the hyperprior for $\alpha$ to 
estimate $\text{E}(\sum_{l=1}^{L} w_{l})$ for any value of the truncation level. The 
Appendix B.2 includes details of the MCMC algorithm for the DDP mixture model 
developed in Section \ref{sec:ddp}, which is a more general version of the density regression 
model of Section \ref{DPM_for_MRL}.

Posterior inference for the density, survival, and mean regression functions can be 
obtained by evaluating $f(t \mid {\boldsymbol x}, G)$, $S(t \mid {\boldsymbol x}, G)$, and 
$\text{E}(T \mid {\boldsymbol x}, G)$ under model (\ref{eqn:dpmm}). Computing proceeds with 
the posterior samples for $G_{L}$, thus involving finite sums at the inference stage. 
Posterior samples for the MRL regression 
function can be efficiently computed using expression (\ref{eqn:dpmm_mrl}), provided the kernel 
MRL function can be readily computed. This is the case for the gamma kernel distribution 
whose MRL function can be expressed in terms of the Gamma function, $\Gamma(a)$, and the gamma 
distribution survival function, $S_{\Gamma}(t)$ \citep{govilt-aggarwal:1983}. More specifically, 
under the gamma density parameterization given in Section \ref{DPM_kernel}, 
$$
m(t \mid \eta,\phi) \, = \, \frac{ t^{e^{\eta}} \, \exp(- e^{\phi} t) \, \exp\{ \phi (e^{\eta} - 1) \} }
{ \Gamma(e^{\eta}) \, S_{\Gamma}(t \mid \eta,\phi) } \, + \, \exp(\eta - \phi) \, - \, t.
$$ 
This expression suffices for the model built from independent kernel components for the survival 
response and covariates, and it can be easily extended to accommodate a gamma kernel density that 
depends on covariates.

The Appendix A includes two synthetic data examples to illustrate the DP mixture 
density regression model. The two examples correspond 
to different scenarios regarding the underlying stochastic mechanism that generates the data.
The simulation truth in the first example involves a finite mixture for the joint response-covariate 
distribution, specified such that the MRL function takes on various non-standard shapes at 
different covariate values. The simulation truth for the second example corresponds to a 
structured parametric setting, where the survival responses are generated from a three-parameter 
extension of the Weibull distribution with its parameters defined through specific regression 
functions.

\section{Dependent DP mixture model for MRL regression}
\label{sec:ddp}

\subsection{The DDP mixture model formulation}
\label{subsec:ddpform}

Often in biomedical studies, researchers are interested in modeling survival times of patients 
under treatment and control groups, where one may expect that the survival distributions for the 
two groups exhibit some similarity. 
%Since the underlying population pre treatment is typically the same, it is reasonable to 
%expect that the survival distributions of the two groups exhibit similarities.  
Modeling groups jointly is thus a natural choice, offering potential learning for the extent of 
similarity, as well as borrowing inferential strength across groups.  We do so by 
generalizing the DP mixture model of Section \ref{sec:curve_reg} to a dependent DP (DDP) mixture 
model. Since we build on the model structure of DP mixture density regression, we achieve 
non-standard MRL regression function shapes, which may also differ across groups contingent 
on the strength of the dependence across experimental groups.

Let $s \in S$ represent in general the index of dependence.  In our case, this indicates the 
experimental group, that is, $S=\{T,C\}$, where $\{ T \}$ is the treatment and $\{ C \}$ is 
the control group.  The model in (\ref{eqn:dpmm}) can be extended to 
$f(t, {\boldsymbol x}\mid G_{s}) =$
$\int k(t,{\boldsymbol x}\mid {\boldsymbol \theta}) \, \text{d}G_s({\boldsymbol \theta})$, 
for  $s \in S$, where we now need a nonparametric prior for the group-specific mixing 
distributions $\{G_s : s \in S\}$. 
%
%We seek to model the distributions in such a way as to incorporate dependencies across experimental 
%groups, while maintaining marginally the DP prior, $G_s \sim \text{DP}$, for each 
%$s\in S$. \citet{maceachern:2000} develops the dependent DP prior in generality with both the weights 
%and atoms under the stick-breaking definition dependent on experimental group: $G_s =$ 
%$\sum_{l=1}^\infty \omega_{ls}\delta_{{\boldsymbol \theta}_{ls}}$. Marginally, 
%$G_s \sim \text{DP}(\alpha_s, G_{0s})$ for each $s \in S$. \citet{maceachern:2000} goes on to 
%describe the computational difficulties in modeling dependencies in the weights across groups, 
%thus motivating development of the common weights model.  In this model, the weights do not change 
%over the groups, only the locations vary, $G_s=$ 
%$\sum_{l=1}^\infty \omega_{l}\delta_{{\boldsymbol \theta}_{ls}}$.  Applications of common weights 
%DDP models include \cite{deiorio-etal:2004}, \cite{rodriguez-horst:2008}, 
%\cite{deiorio-etal:2009}, \citet{kottas-behseta:2012}, and \cite{kottas-fronczyk:2014}.  
%
The DDP prior structure allows for dependence across experimental groups, while maintaining the 
DP prior marginally for each $G_s$. The DDP prior builds from the DP constructive definition, and,
in its most general form, extends the structure for $G$ to $G_s =$ 
$\sum_{l=1}^\infty w_{ls} \, \delta_{{\boldsymbol \theta}_{ls}}$, for $s \in S$.
Typically, simplified versions are employed with common weights or common atoms. 
Here, we opt for a common atoms DDP prior model, $G_s =$ 
$\sum_{l=1}^\infty w_{ls} \, \delta_{{\boldsymbol \theta}_l}$, with 
$\boldsymbol{\theta}_l \stackrel{\text{i.i.d.}}{\sim} G_{0}$, as in the DP prior, and 
group-specific dependent weights defined through a bivariate beta distribution for the 
latent stick-breaking variables. The construction for the weights is detailed later in 
the section.

%
%While computationally convenient and a useful extension of the basic DP prior, assuming the 
%same weights has potential disadvantages in our setting.  A practical disadvantage of the common 
%weights DDP construction involves applications with a moderate to large number of covariates.  
%For such cases, the common weights prior requires building dependence across $s\in S$ for a 
%large number of kernel parameters, whereas modeling dependence through the weights is not 
%affected by the dimensionality of the mixture kernel
%In situations where we might expect similar locations across groups, modeling dependence 
%through the weights is more attractive.  In our context, we may expect the two groups to 
%be comprised of similar components which however exhibit different prevalence across 
%survival time. 
%

A practical advantage of modeling dependence only through the weights is that the construction
is not affected by the dimension of the mixture kernel parameter vector, i.e., by the number
of covariates. Moreover, the common atoms prior model structure is useful if we expect the 
group-specific distributions to be comprised of similar components which however exhibit 
different prevalence across survival time.

Hence, the DDP mixture model can be expressed as 
\begin{equation}
f(t, {\boldsymbol x}\mid G_{s}) \, = \, 
\int k(t,{\boldsymbol x}\mid {\boldsymbol \theta}) \, \text{d}G_s({\boldsymbol \theta}) 
\, = \, \sum_{l = 1}^\infty w_{ls} \, k(t, {\boldsymbol x} \mid {\boldsymbol \theta}_l),
\,\,\,\,\, s \in \{T,C\}.
\label{eqn:ddpmm}
\end{equation}
%where $G_s \sim \text{DDP}({\boldsymbol \Phi}, G_0)$, with ${\boldsymbol \Phi}$ representing 
%the parameters associated with the construction of the group-specific dependent weights. 
Then, the conditional response density becomes
$f(t \mid {\boldsymbol x},  G_s) =$ 
$ \sum_{l=1}^\infty  q_{ls}({\boldsymbol x}; {\boldsymbol \theta}_l) \, 
k(t\mid {\boldsymbol x},{\boldsymbol \theta}_l)$, and the conditional survival function 
can be written as
\begin{eqnarray}
\label{eqn:ddpsurv}
S(t \mid {\boldsymbol x}, G_s) &=& 
\sum_{l=1}^\infty  q_{ls}({\boldsymbol x}; {\boldsymbol \theta}_l) \, 
S(t \mid {\boldsymbol x}, {\boldsymbol \theta}_l)
\end{eqnarray}      
where $q_{ls}({\boldsymbol x}; {\boldsymbol \theta}_l) =$
$w_{ls} \, k({\boldsymbol x} \mid {\boldsymbol \theta}_l)/
\{\sum_{r=1}^\infty  w_{rs} \, k({\boldsymbol x} \mid {\boldsymbol \theta}_r)\}$. Likewise, 
the group-specific mean regression function is $\text{E}(t\mid{\boldsymbol x}, G_s) =$
$\sum_{l=1}^\infty q_{ls}({\boldsymbol x}; {\boldsymbol \theta}_l) \,
\text{E}(t\mid{\boldsymbol  x},{\boldsymbol \theta}_l)$. 
We thus recognize again the local weighted mixture structure with covariate-dependent weights,
which now change also with the treatment (group) assignment.

Using (\ref{eqn:ddpsurv}) with definition (\ref{eqn:mrl}), the MRL regression function of the 
general model becomes
\begin{eqnarray}
m(t\mid {\boldsymbol x}, G_s) \, = \,
\frac{\int_t^\infty S(u\mid {\boldsymbol x},G_s) \, \text{d}u}{S(t\mid {\boldsymbol x},G_s)} 
\, = \, \sum_{l=1}^\infty q^*_{ls}(t,{\boldsymbol x} ;{\boldsymbol \theta}_l) \, 
m(t\mid {\boldsymbol x},{\boldsymbol \theta}_l)
\label{eqn:ddpmm_mrl}
\end{eqnarray}
where the weights can be written as
\begin{equation}
q^*_{ls}(t, {\boldsymbol x} ; {\boldsymbol \theta}_l) \, = \,
\frac{w_{ls} \, k({\boldsymbol x}\mid {\boldsymbol \theta}_l) \,
S(t\mid{\boldsymbol x}, {\boldsymbol \theta}_l)}
{\sum_{r=1}^{\infty} w_{rs} \, k({\boldsymbol x}\mid {\boldsymbol \theta}_r) \,
S(t\mid{\boldsymbol x},{\boldsymbol \theta}_r)}.
\label{DDP_MRL_weights}
\end{equation}
Again, the local weighted mixture formulation extends to the MRL regression, with the local 
adjustment over the covariates, time, and (now) groups controlled by separate components 
of the mixture weights. Importantly, the model allows flexible MRL functions that can vary
in shape across treatment groups at the same covariate values or at the same time points.
The practical utility of this model feature is demonstrated in Section \ref{sec:lung_cov}
with relevant results from the analysis of the small cell lung cancer data.

Next, we turn to the construction of the dependent weights of $G_s$. 
The stick-breaking method to construct the weights uses latent variables
$v_l \stackrel{\text{i.i.d.}}{\sim} \text{Beta}(1,\alpha)$, but one can equivalently work 
with $\zeta_l \stackrel{\text{i.i.d.}}{\sim} \text{Beta}(\alpha, 1)$ adjusting accordingly 
the expression for the weights. Therefore, the weights of the DDP prior for 
$G_s =$ $\sum_{l=1}^\infty w_{ls} \, \delta_{{\boldsymbol \theta}_l}$ are:
$$
w_{1s} \, = \, 1 - \zeta_{1s}, \,\,\,\,\,
w_{ls} \, = \, (1- \zeta_{ls}) \, \prod\nolimits_{r=1}^{l-1} \zeta_{rs}, \,\,\, l \geq 2,
$$
with $(\zeta_{lC},\zeta_{lT})$ i.i.d. from a bivariate beta distribution for which the 
marginals are $\text{Beta}(\alpha,1)$ distributed. This construction incorporates dependence
between the two groups (additional to the dependence induced by the common atoms), while 
maintaining the $\text{DP}(\alpha, G_{0})$ prior marginally for the group-specific mixing 
distribution $G_s$.

We work with a bivariate beta distribution from \citet{nadarajah-kotz:2005}, defined 
constructively through products of independent beta distributed random variables. 
In particular, to define the bivariate beta distribution for $(X,Y)$, start with independent 
random variables, $U\sim \text{Beta}(a_1,b_1)$, $V\sim \text{Beta}(a_2,b_2)$, and 
$W\sim \text{Beta}(c,b)$, subject to the constraint, $c=$ $a_1 + b_1 =$ $a_2 + b_2$. 
Then, define $X = U \, W$ and $Y = V \, W$. The marginals are given by 
$X \sim \text{Beta}(a_1, b_1 + b)$ and $Y\sim \text{Beta}(a_2, b_2 +b)$. We can thus obtain 
the desired beta marginals for $\zeta_{l C}$ and $\zeta_{l T}$ by setting $b_1 =$ 
$b_2 \equiv$ $1 - b$, and $a_{1}=a_{2} \equiv \alpha$.
The DDP prior model is completed with hyperpriors for $\alpha > 0$ and $b \in (0,1)$, 
as well as for parameters of the baseline distribution $G_0$ for the common atoms.

For MCMC posterior simulation, we work with the $(U_l, V_l, W_l)$ to update 
$(\zeta_{lC},\zeta_{lT})$, as well as hyperparameters $\alpha$ and $b$, thus bypassing 
the complex form of the bivariate beta density for the $(\zeta_{lC},\zeta_{lT})$.
The Appendix B.2 details the blocked Gibbs sampler for the DDP mixture model.

The latent (independent) random variables $(U_l, V_l, W_l)$ can also be used to study 
properties of the bivariate beta distribution. In particular, working with product moments
of the bivariate beta distribution for the $(\zeta_{lC},\zeta_{lT})$, we can build the 
correlation of $(w_{lC},w_{lT})$, and then the correlation of $(G_{C}(B),G_{T}(B))$ (for any 
specified measurable set $B$). The correlation between the group-specific survival times, $T_{s}$, 
$s \in \{T,C\}$, can also be obtained. Details are provided in the Appendix B.1.

\subsection{Connections with DDP survival regression models}
\label{other_DDP_models}

The model developed in Section \ref{subsec:ddpform} combines ideas from DP mixture density 
regression and covariate-dependent DDP mixture models, the two main NPB approaches to 
fully nonparametric regression. Here, we contrast the MRL regression formulation implied by 
our model with the one arising from general DDP regression, and from the more structured
DDP model in \cite{deiorio-etal:2009}, a standard reference on DDP survival regression.

For simpler exposition, we consider the setting with one continuous covariate, $x$, and 
one binary covariate, $s \in \{0,1\}$ (say, a treatment indicator), as in the small cell 
lung cancer data set of Section \ref{sec:lung_data}. Under the DDP mixture approach, 
the prior model is defined directly for the conditional response density:
$$
f(t \mid G_{s,x}) \, = \, 
\int k(t \mid {\boldsymbol \varphi}) \, \text{d}G_{s,x}({\boldsymbol \varphi}) 
\, = \, \sum_{l = 1}^\infty \omega_{ls}(x) \, k(t \mid {\boldsymbol \varphi}_{ls}(x)),
$$
with $k(t \mid {\boldsymbol \varphi})$ a kernel density on $\mathbb{R}^{+}$, under which 
the MRL function of the mixture distribution is well defined. Here, the mixture weights, 
$\omega_{ls}(x)$, and atoms, ${\boldsymbol \varphi}_{ls}(x)$, depend on both covariates, 
and the notation takes into account the discrete (binary) nature of $s$. Then the 
covariate-dependent MRL function can be expressed as 
$$
m(t \mid G_{s,x}) \, = \,  \sum_{l=1}^{\infty}
\left\{
\frac{ \omega_{ls}(x) \, S(t \mid {\boldsymbol \varphi}_{ls}(x))}
{\sum_{r=1}^{\infty} \omega_{rs}(x) \, S(t \mid {\boldsymbol \varphi}_{rs}(x))} \right\}
\, m(t \mid {\boldsymbol \varphi}_{ls}(x)),
$$
where $S(t \mid {\boldsymbol \varphi})$ and $m(t \mid {\boldsymbol \varphi})$ are the kernel 
survival and MRL function, respectively. Hence, the prior model structure is generally similar 
to the one in (\ref{eqn:ddpmm_mrl}), although the fashion in which the different weight components 
control local adjustment over time and the covariates is not as clear as in (\ref{DDP_MRL_weights}).

The general DDP prior formulation requires group-specific stochastic processes (indexed by the 
continuous covariate values) for the atoms ${\boldsymbol \varphi}_{ls}(x)$, and another set 
of stochastic processes in $x$ for the covariate-dependent stick-breaking variables that define 
the weights $\omega_{ls}(x)$. To retain the DP prior marginally at any $(s,x)$, the 
latter stochastic processes must have beta marginals. Evidently, a general DDP prior model 
becomes more challenging to formulate as the dimension of the covariate space increases, and 
it is also difficult to estimate with small to moderate amounts of data.

A substantial simplification of the general DDP structure is offered by a linear-DDP formulation, 
under which only the atoms depend on the covariates through (transformations of) linear 
regression functions. Consider a two-parameter mixture kernel, $k(t \mid h(\theta),\xi)$, where
$\xi$ is a scale parameter, and $h(\theta)$ controls the center of the distribution (say, the 
mean or median parameter), with $\theta \in \mathbb{R}$ and $h$ a specified link function. 
Then, a possible linear-DDP mixture model for the conditional response density can be written as:
$$
f(t \mid G_{s,x}) \, = \, 
\sum_{l = 1}^\infty \omega_{l} \, k(t \mid h(\theta_{l}(s,x)), \xi_{l}),
$$
where $\theta_{l}(s,x) =$ $\beta_{0l} + \beta_{1l} \, s + \beta_{2l} \, x$.
Examples for the kernel include the gamma density with mean $\exp(\theta)$, and the 
log-normal density with median $\exp(\theta)$ and $\xi$ corresponding to the standard deviation 
of the normal whose transformation yields the log-normal. The latter choice essentially 
yields the linear-DDP regression model in \cite{deiorio-etal:2009}.

The MRL regression under the linear-DDP model can be written as 
$$
m(t \mid G_{s,x}) \, = \,  \sum_{l=1}^{\infty}
\left\{ \frac{ \omega_{l} \, S(t \mid h(\theta_{l}(s,x)),\xi_{l})}
{\sum_{r=1}^{\infty} \omega_{r} \, S(t \mid h(\theta_{r}(s,x)),\xi_{r})} \right\} 
\, m(t \mid h(\theta_{l}(s,x)), \xi_{l})
$$
where $S(t \mid h(\theta),\xi)$ and $m(t \mid h(\theta),\xi)$ are the kernel survival and 
MRL function, respectively. The linear-DDP structure is restrictive in terms of the 
implied prior for the MRL function. In contrast to the form in (\ref{DDP_MRL_weights}),
here there is a single function (the kernel survival function) that defines the 
dependence of the weights on time and the covariates. Lacking the local adjustment over
values in the covariate space impedes model flexibility. For instance, under the 
log-normal kernel, the mean regression function for the control group ($s=0$)
and treatment group ($s=1$) is 
$\sum_{l = 1}^\infty \omega_{l} \, \exp\{ (0.5 \xi_{l}^{2} + \beta_{0l}) + \beta_{2l} x \}$
and $\sum_{l = 1}^\infty \omega_{l} \, 
\exp\{ (0.5 \xi_{l}^{2} + \beta_{0l} + \beta_{1l}) + \beta_{2l} x \}$. 
Hence, the model can not uncover general non-linear, non-monotonic regression relationships, 
and furthermore the mean regression function varies across groups in a rigid fashion.

\subsection{Synthetic data examples}
\label{simulation_DDPM}

In this section, we consider two synthetic data examples to investigate the performance 
of the DDP mixture model without covariates, that is, 
$f(t \mid G_{s}) =$ $\sum_{l = 1}^\infty w_{ls} \, k(t \mid \eta_l,\phi_l)$, where 
$k(t \mid \eta,\phi) \propto$ $t^{e^\eta -1}\text{exp}(-e^\phi t)$, and the baseline distribution 
is $G_{0}(\eta, \phi)=$ $\text{N}_2((\eta, \phi)\mid{\boldsymbol \mu}, {\boldsymbol \Sigma})$.
The model is completed with the following hyperpriors: 
${\boldsymbol \mu} \sim \text{N}_2(\boldsymbol{\mu}\mid a_\mu, B_\mu)$,
${\boldsymbol \Sigma} \sim  \text{IWish}({\boldsymbol \Sigma}\mid a_\Sigma, B_\Sigma)$, 
$\alpha \sim \Gamma(\alpha \mid a_\alpha, b_\alpha)$ (gamma distribution with mean
$a_\alpha / b_\alpha$), and $b \sim \text{Unif}(b\mid 0,1)$.

The first simulation scenario involves distributions defined by mixtures of 
Weibull distributions with the same atoms and different weights. We expect the DDP mixture model 
to perform well under this scenario since the truth shares the same structure as the DDP prior 
model. The true density, survival, and MRL functions are shown in Figure \ref{fig:DDPMMsim1fun}.
Note that the two groups have bimodal densities with modes at essentially the same locations, but 
with differing prevalence. The second scenario is also based on mixtures of Weibull distributions, 
but, in this case, with different weights and atoms. The intention is to test the model's inferential 
ability for distributions with quite different features. In particular, the second group density
exhibits a single mode in between the two modes of the first group density; see Figure 
\ref{fig:DDPMsim2post} for the true density, survival, and MRL functions.

\subsubsection{Simulation 1}

We work with Weibull mixture distributions that share atoms, but have different weights:  
$T_1 \sim 0.7\text{Weib}(2,8) + 0.1\text{Weib}(3,10)  + 0.05\text{Weib}(4,30) + 0.15\text{Weib}(8,40)$ and 
$T_2 \sim 0.5\text{Weib}(2,8) + 0.05\text{Weib}(3,10)  + 0.025\text{Weib}(4,30) + 0.425\text{Weib}(8,40)$.  
We sample $250$ survival times from the first population and $100$ survival times from the second. 
%We do not consider censoring or covariates here. 
%We place a gamma prior on $\alpha$ with shape parameter $2$ and rate parameter $0.8$.  
%The number of components is conservatively set at $40$.  
Regarding the hyperparameters, we set: $a_\alpha = 2$, $b_\alpha = 0.8$, $a_\mu = (1.87, 0.25)'$, 
$B_\mu =$ $B_\Sigma=((0.27, 0)',(0, 0.27)')$, and $a_\Sigma = 4$.  
%After burn in and thinning, we obtain 2000 independent posterior samples. 

%\begin{comment}
\begin{figure}[t!]
\centering
    \includegraphics[width=0.3\textwidth]{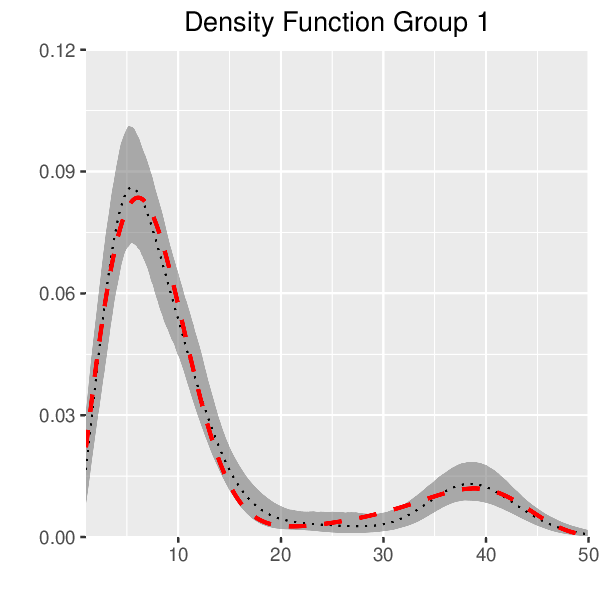}
    \includegraphics[width=0.3\textwidth]{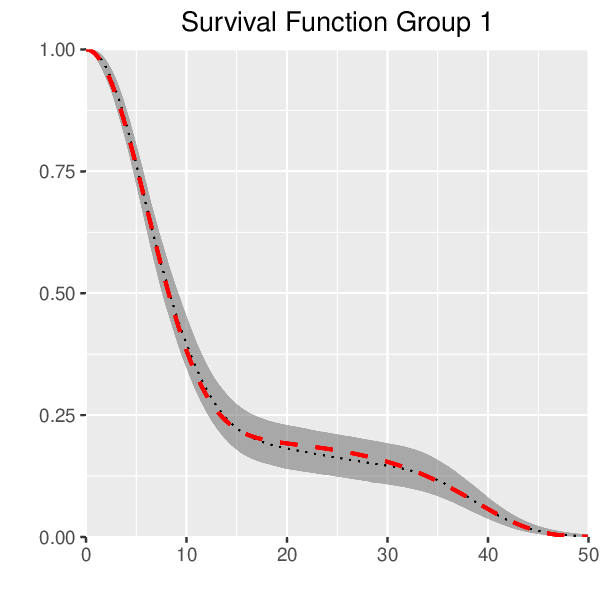} 
    \includegraphics[width=0.3\textwidth]{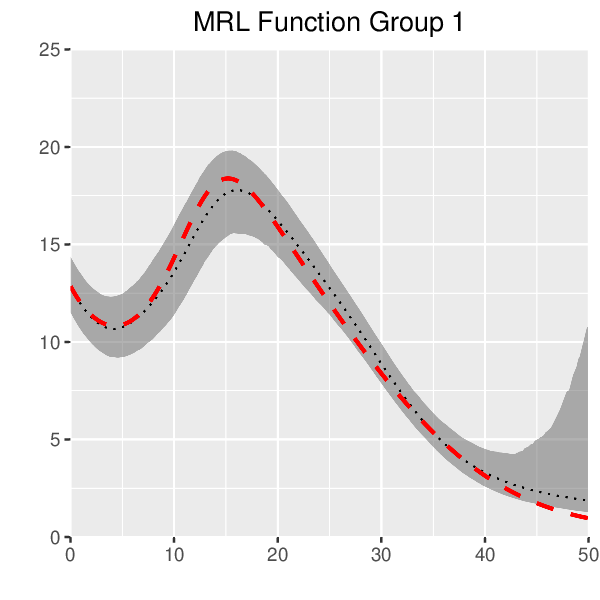}
     \includegraphics[width=0.3\textwidth]{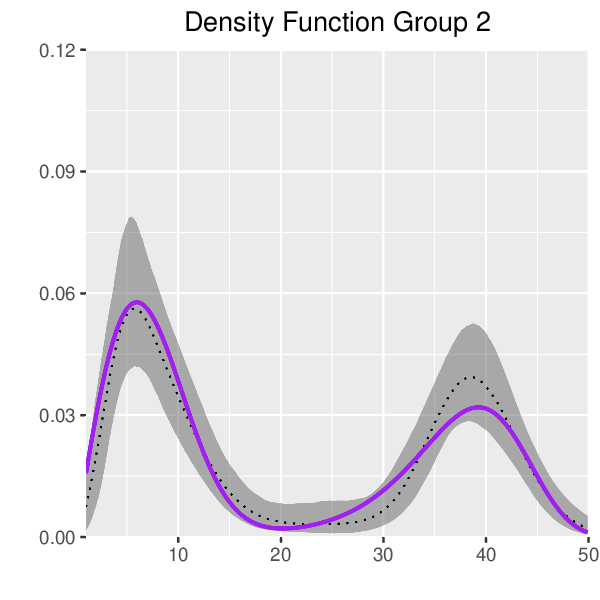} 
     \includegraphics[width=0.3\textwidth]{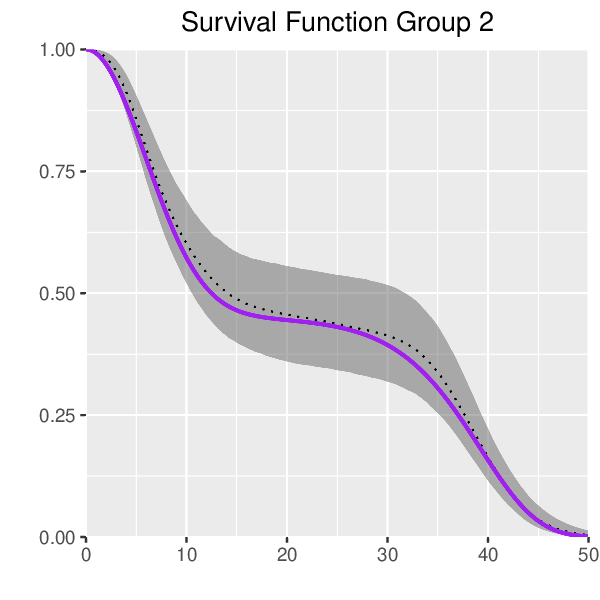}
      \includegraphics[width=0.3\textwidth]{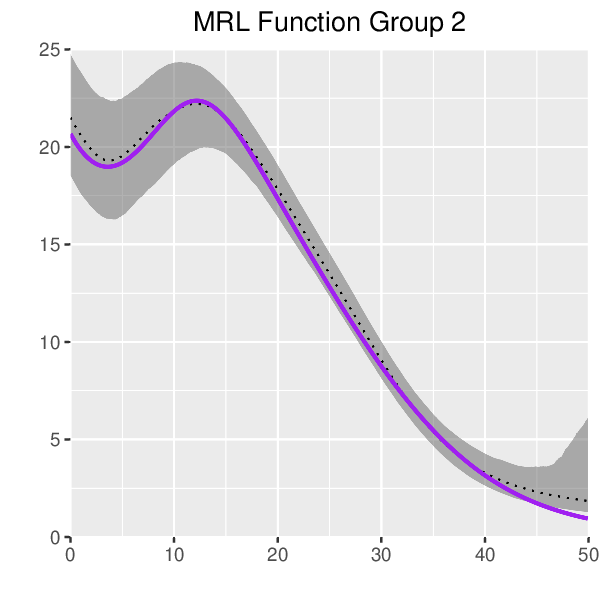}
\caption{Simulation 1. Posterior point and $95\%$ interval estimates for the 
density (left), survival (middle), and MRL (right) functions. The corresponding 
true functions are denoted by the dashed red lines (Group 1) and solid purple 
lines (Group 2).}
        \label{fig:DDPMMsim1fun}
\end{figure}
%\end{comment}

%
%The $95\%$ posterior credible intervals for $\alpha$, $b$, and $\text{Corr}(G_C, G_T)$ 
%are given by $(1.89, 14.45)$, $(0.15, 0.78)$, and $(0.59, 0.88)$, respectively.  
%The model is able to express the features of the functionals, and the true population density 
%is captured within the $95\%$ interval estimates save for the very tail where data is very sparse.  
%In particular, the flexibility of the model is demonstrated in the MRL function.  The true MRL 
%is non-standard in both groups: initially decreasing, followed by an increase after about 
%time $5$, and then decreasing again after about time $12$.  The difference in sample size 
%between the two groups is indicated by the slightly larger interval bands in Group 2 for 
%the majority of the support of the data. 
%
Posterior point and interval estimates for the density, survival, and MRL functions are provided 
in Figure \ref{fig:DDPMMsim1fun}. Despite the moderate sample sizes, the model performs well in 
inference for all functionals, including successfully recovering the non-standard shape of the 
group-specific MRL functions. As expected from the difference in the sample size for the two 
groups, the posterior interval bands are wider for the second group estimates.

\subsubsection{Simulation 2}

Here, the Weibull mixtures have different weights, atoms, and number of components: 
$T_1 \sim$ $0.5 \, \text{Weib}(2,4) +$ $0.05 \, \text{Weib}(0.6,4) +$ 
$0.025 \, \text{Weib}(5,15) +$ $0.425 \, \text{Weib}(8,30)$, and $T_2 \sim$
$0.02 \, \text{Weib}(0.6,1) +$ $0.02 \, \text{Weib}(2,4) +$ $0.66 \, \text{Weib}(5,15) +$
$0.2 \, \text{Weib}(2,8) +$ $0.1 \, \text{Weib}(4,30)$. 
We generate $250$ survival times from each population.  
%All observations are fully observed, and no covariates are considered. Once again, we use 
%a uniform prior on $b$, and  gamma prior on $\alpha$ with shape parameter $2$ and rate 
%parameter $0.8$.  The number of components is set at $40$, which is a conservative value 
%for these data.  
%We assume $a_\mu =(3.02,  0.54)'$, $B_\mu=B_\Sigma=((0.1,0)',(0,0.1)')$, and $a_\Sigma = 4$.   
%After burn in and thinning, we obtain $2000$ independent posterior samples.  
The hyperparameters are specified as follows: $a_\alpha = 2$, $b_\alpha = 0.8$, 
$a_\mu =(3.02, 0.54)'$, $B_\mu=$ $B_\Sigma=((0.1,0)',(0,0.1)')$, and $a_\Sigma = 4$.

%
%The $95\%$ posterior credible intervals for $\alpha$, $b$, and $\text{Corr}(G_C, G_T)$ 
%are given by $(0.76, 3.88)$, $(0.12, 0.72)$, and $(0.62, 0.84)$, respectively. 
%The posterior results for the density, survival, and MRL functions are shown 
%in Figure \ref{fig:DDPMsim2post}.  Despite the difference in the features of the functionals 
%between the two groups, the model is able to capture the features of each group with accuracy.  
%This is especially exciting for the MRL functions.  The MRL functions are quite different from 
%one another, and both are non-standard shapes.  The model has no problem capturing both shapes 
%of the MRL functions. The only area where we can see struggle in the model for the MRL function 
%inference is in the tails of the functionals.  The true MRL function of Group 1 is slightly above 
%the upper interval estimate of the model.  This may be just due to the random nature of simulated 
%data; this simulated data may suggest a lower MRL function in the tail.  Another possibility is the 
%extreme difference between the MRL functions of the two groups in the tails.  Group 1 shoots up 
%sharply, while Group 2 remains gradually decreasing.  A third contributor to the tail struggle is 
%that the sparsity of the data in this area, so models in general a have a tougher time achieving 
%accuracy.  Even with these elements against the model, the struggle is not significant.  
%

%\begin{comment}   
\begin{figure}[t!]
\centering
    \includegraphics[width=0.3\textwidth]{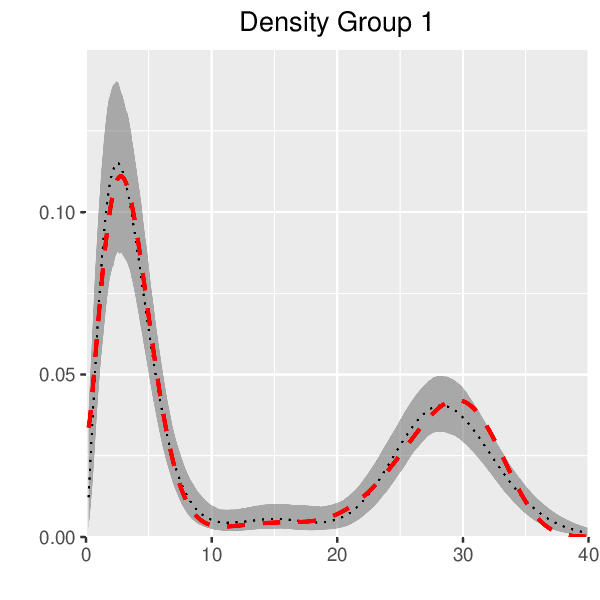}
    \includegraphics[width=0.3\textwidth]{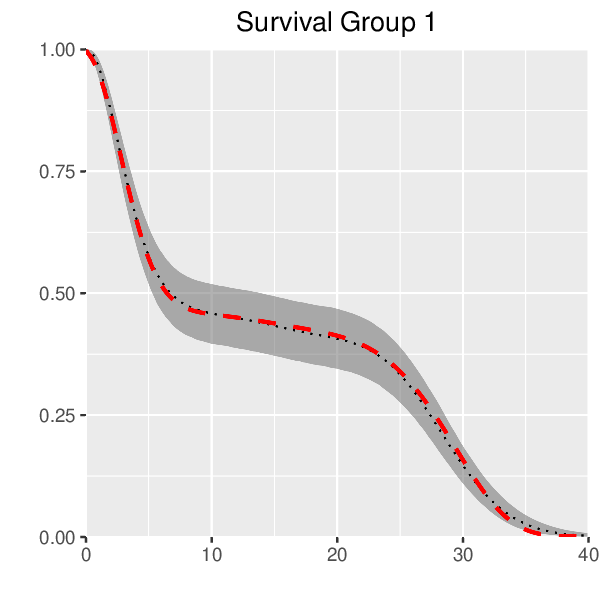}
    \includegraphics[width=0.3\textwidth]{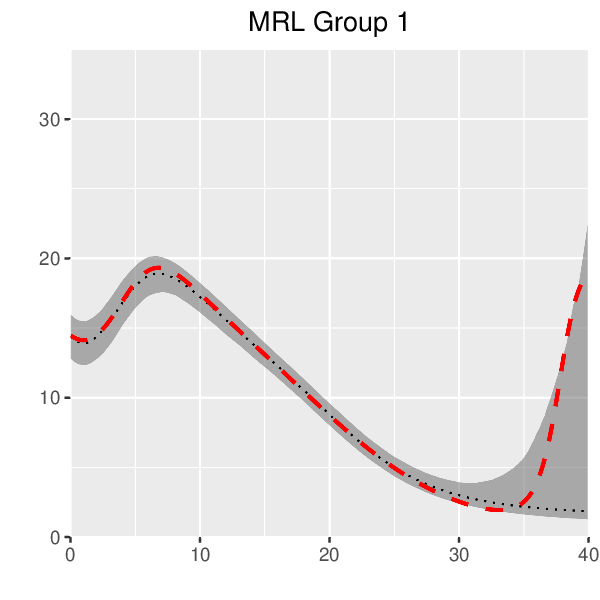}
    \includegraphics[width=0.3\textwidth]{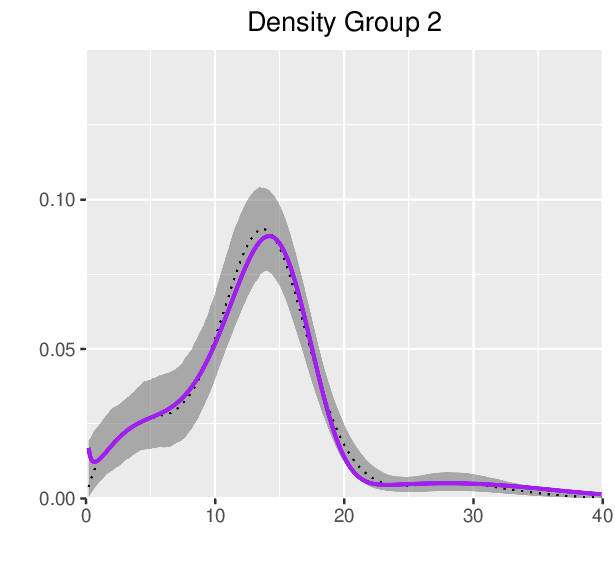}
    \includegraphics[width=0.3\textwidth]{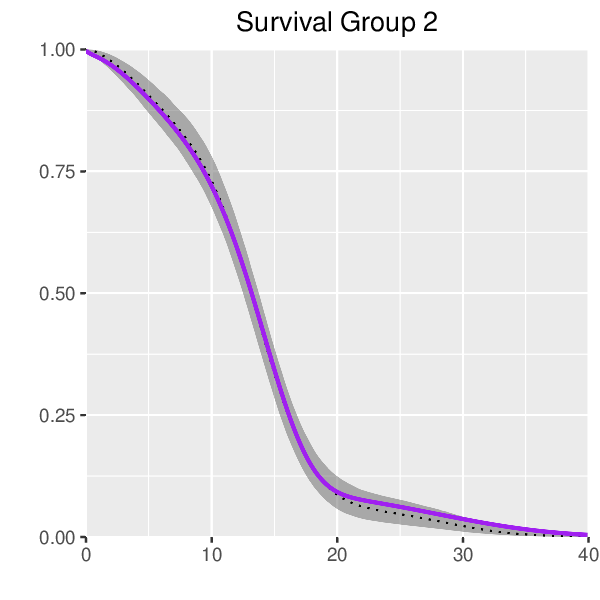}
    \includegraphics[width=0.3\textwidth]{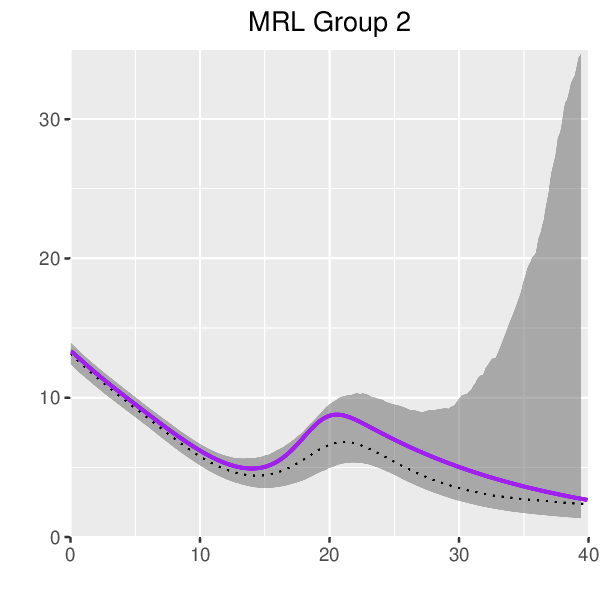}
\caption{Simulation 2. Posterior point and $95\%$ interval estimates for the 
density (left), survival (middle), and MRL (right) functions. The corresponding 
true functions are denoted by the dashed red lines (Group 1) and solid purple 
lines (Group 2).}\label{fig:DDPMsim2post}
\end{figure}
%\end{comment}

As shown in Figure \ref{fig:DDPMsim2post}, the model is successfully estimating the group-specific 
density, survival, and MRL functions, even though the simulation truth is now different from the 
model structure. In particular, the MRL function has non-standard shape which changes in a 
non-trivial fashion across the two groups. The first group MRL function depicts a rapid increase 
for survival times at the tail of the corresponding density. Regarding the proportion of data 
in the tail, only one out of the 250 simulated survival times has value greater than $35$.
The MRL function point estimate does not capture the increase in that range of survival times, 
but the $95\%$ posterior uncertainty bands essentially include the true function.

%
%The results from the two simulations demonstrate the practical utility of the DDP mixture 
%model. The model is able to incorporate dependence across two populations to achieve accurate 
%inference in the functionals of each population. In particular, the model provides flexible 
%MRL inference for two groups that exhibit MRL functions with different features across the 
%range of survival.
%

\section{Small cell lung cancer data example}
\label{sec:lung_data}

We consider a dataset, provided in \cite{ying-etal:1995}, that comprises survival times, 
in days, of patients with small cell lung cancer. 
The data has been used as a test case for classical and Bayesian semiparametric regression models, 
typically, with a linear regression for a central feature (e.g., median) of the log-transformed 
survival times; see, e.g., \cite{kottas-krnjajic:2009} and further references therein. 
The patients were randomly assigned to one of two treatments referred to as Arm A and Arm B. 
Arm A patients received cisplatin (P) followed by etoposide (E), while Arm B patients 
received E followed by P. The Arm A group consists of $62$ survival times, $15$ of which 
are right censored, whereas the Arm B group consists of $59$ survival times, $8$ of which are 
right censored. The age of each patient at entry in the clinical study is also available. 
We begin the data analysis in Section~\ref{sec:lung_nocov} working with treatment as the 
only covariate, and incorporate the age covariate in Section~\ref{sec:lung_cov}.

\subsection{Comparative inference for the treatment groups}
\label{sec:lung_nocov}

\subsubsection{Results under the DDP mixture model}

\begin{figure}[t!]
 \centering
    \includegraphics[width=0.45\textwidth]{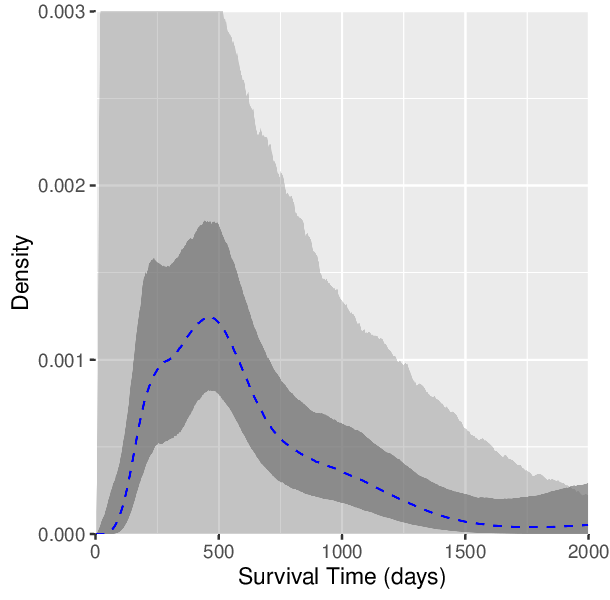}
    \includegraphics[width=0.45\textwidth]{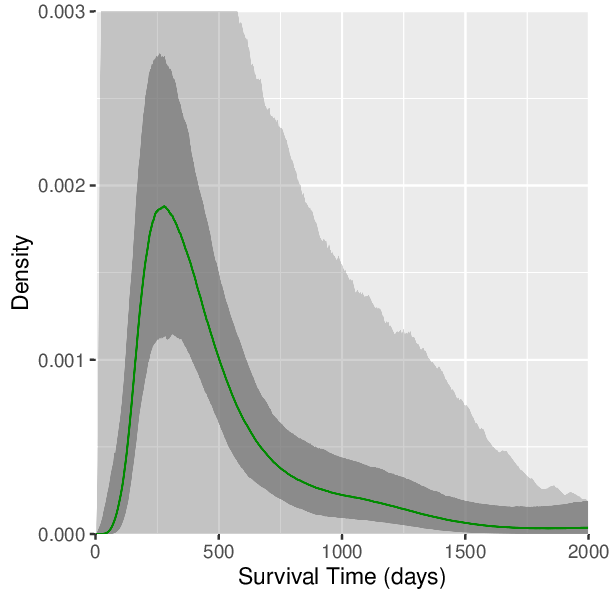}
    \includegraphics[width=0.45\textwidth]{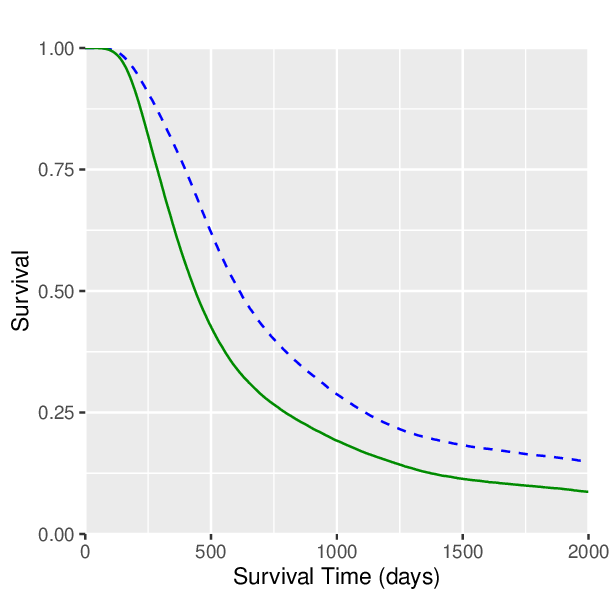}
    \includegraphics[width=0.45\textwidth]{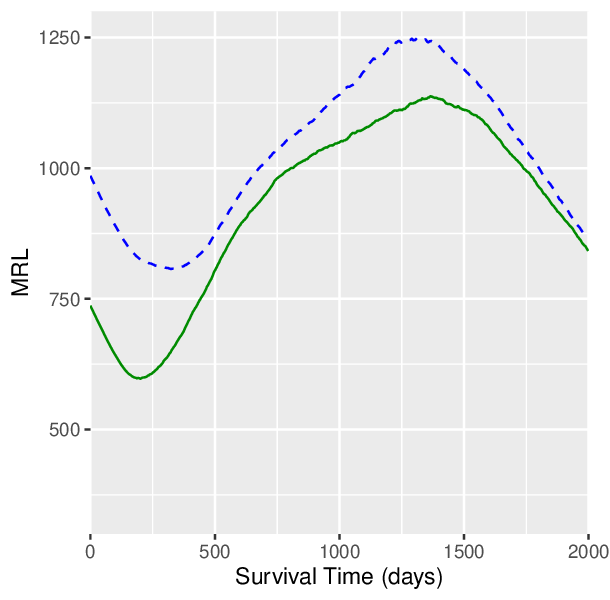} 
\caption{Small cell lung cancer data. Top panels: prior $95\%$ interval estimates (light grey bands), 
and posterior point and $95\%$ interval estimates (dark grey bands) of the density function for Arm A 
(top left panel) and Arm B (top right panel). Note that the upper bound for the prior interval estimates
extends to density values above $0.009$ at small values for survival time. 
Bottom panels: posterior point estimate of the survival function (bottom left panel) and the mean 
residual life function (bottom right panel) for Arm A (blue dashed lines) and Arm B (green solid lines).}
\label{fig:DDPMMabMRL}
\end{figure}

We fit the gamma DDP mixture model, given in Section \ref{simulation_DDPM}, such that $s$
indexes the Arm A and Arm B treatment groups. 
For the hyperparameters, we set: $a_\alpha = 2$, $b_\alpha = 0.5$, $a_\mu = (3.0,-2.8)$, 
$B_\mu = B_\Sigma = ((0.19, 0)',(0,0.19)')$, and $a_\Sigma = 4$.  
The effect of any particular prior choice can be usefully studied through the implied 
prior uncertainty for the survival density (or other functionals). The prior $95\%$ interval
bands for the Arm A and Arm B density are plotted in Figure \ref{fig:DDPMMabMRL}.

%\end{comment}

Inference results for the density, survival, and MRL functions are shown in Figure \ref{fig:DDPMMabMRL}. 
The model estimates heavy tailed densities, with the mode at about 450 days and 350 days for Arm A 
and Arm B, respectively. Comparing the prior and posterior $95\%$ credible interval bands suggests
substantial learning from the data. The survival function point estimates indicate that Arm A has 
higher survival rate starting from about 200 days. The estimated treatment-specific MRL functions 
exhibit a non-linear, non-monotonic trend with Arm A having higher mean residual life over the 
effective range of survival times.  

\begin{figure}[t!]
\centering
\includegraphics{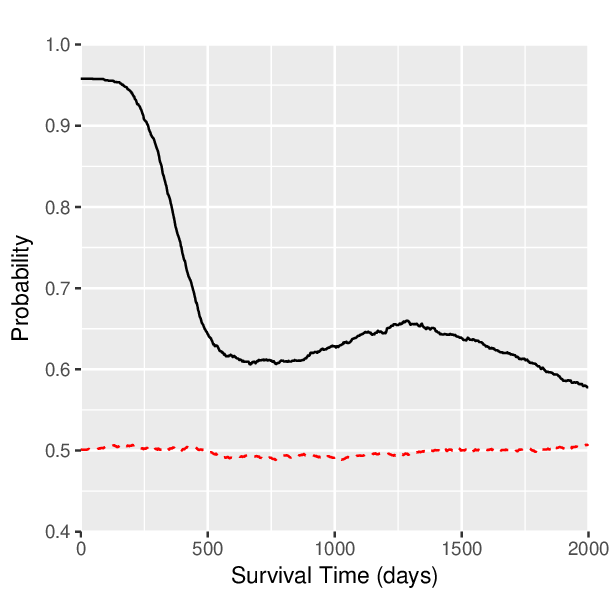}
\caption{Small cell lung cancer data. The posterior probability (black solid line) and prior 
probability (red dashed line) of the MRL function of Arm A being higher than the MRL function 
of Arm B over a grid of survival times (days).}
\label{fig:MRLdiffprobDDP}
\end{figure}

For comparison of the treatment-specific MRL functions that incorporates posterior uncertainty, 
we work with the posterior probability $\text{Pr}(m_A(t) > m_B(t) \mid \text{data})$ over a grid of 
survival times $t$, where $m_A$ and $m_B$ is the MRL function under Arm A and Arm B, respectively. 
The results are reported in Figure \ref{fig:MRLdiffprobDDP}, where we also include the prior 
probability, $\text{Pr}(m_A(t) > m_B(t))$. Note that the prior does not favor either of the 
two treatments. The posterior probability is greater than $0.5$ over the entire effective range 
of survival times; it is fairly large (about $0.95$) for the first 200 days, then it decreases 
to about $0.65$ at around 500 days, from which point on it essentially stabilizes to values that 
range from $0.58$ to $0.65$.

\subsubsection{Model comparison}
\label{sec:cpo}

Although we are not aware of other NPB models that explore inference for MRL regression,
as previously mentioned, the small cell lung cancer data set has been used for illustration 
of semiparametric survival regression models. In particular, \cite{kottas-krnjajic:2009} develop 
a Bayesian semiparametric quantile regression model, based on a linear quantile regression 
function and a non-parametric scale mixture of uniform densities for the error distribution,
and apply it to the log-transformed survival times, using the treatment indicator as the 
only covariate. We compare the predictive performance of the DDP mixture model 
through the criterion used in \cite{kottas-krnjajic:2009}, which is based on conditional 
predictive ordinate (CPO) values.

%
%The CPO of the $i^{th}$ observation, $\text{CPO}_i$, can be expressed in terms of the joint 
%posterior distribution of the model parameters,  ${\boldsymbol \Psi}$, given {\it all} 
%observations: $\text{CPO}_i = \left(\int f(t_i|{\boldsymbol \Psi}, x_i)^{-1} \pi({\boldsymbol \Psi}| %data)d{\boldsymbol \Psi}\right)^{-1}$. The expression often does not have a closed form, 
%so MCMC approximation is used (see, for example, \cite{chen-etal:2000}). The DDP mixture 
%model requires a slightly different expression for the CPO values.  We provide the expression 
%and derivation details in the Supplementary Material.
%

The CPO for an observed survival time $t_i$ is defined by the value of the posterior predictive
density at $t_i$, given the data with $t_i$ excluded. The definition can be extended to right 
censored observations replacing the predictive density by the predictive survival function. 
Large CPO values indicate agreement between the associated observations and the model. Models
can be compared by summarizing the CPOs, say, through the average of the log-CPO values, which 
we refer to as the ALPML criterion. For many hierarchical Bayesian models, CPOs can be computed 
using posterior samples obtained by fitting the model once to all observations; see, e.g., 
\cite{chen-etal:2000}. This is also possible for the DDP mixture model; the details are 
provided in Appendix B.3.

The ALPML value for the DDP mixture model is $-6.05$. The ALPML values reported in 
\cite{kottas-krnjajic:2009} are $-6.91$ for the semiparametric quantile regression 
model, and $-11.56$ for a Weibull proportional hazards model. The semiparametric 
model provides improvement in predictive performance relative to the parametric proportional 
hazards regression model. The fully nonparametric DDP mixture model offers further improvement.

\subsection{Incorporating the age covariate}
\label{sec:lung_cov}

%Here, we incorporate the age (in years) of the subjects, upon entrance into the study, that 
%is also available in the small cell lung cancer dataset.  The researchers did not select 
%subjects from particular ages, so it is not a fixed covariate, and it can thus be incorporated 
%into the model through a joint response-covariate distribution.

Here, we consider the analysis of the data including both available covariates, the 
treatment indicator ($s$) and patient age at entry in the clinical study ($x$). 
We apply the model in (\ref{eqn:ddpmm}) with a product mixture kernel, 
$k(t, x \mid \eta,\phi,\beta,\kappa^{2}) =$ $k(t \mid \eta,\phi) \, k(x \mid \beta,\kappa^{2})$, 
where $k(t \mid \eta,\phi) \propto$ $t^{e^\eta -1}\text{exp}(-e^\phi t)$ and 
$k(x \mid \beta,\kappa^{2})$ is the normal density with mean $\beta$ and variance $\kappa^{2}$.
The baseline distribution $G_0$ comprises independent components: a bivariate normal distribution
$\text{N}_2({\boldsymbol \mu}, {\boldsymbol \Sigma})$ for $(\eta, \phi)$; a normal distribution 
$\text{N}(\lambda, \tau^2)$ for $\beta$; and an inverse-gamma distribution 
$\Gamma^{-1}(a, \rho)$ for $\kappa^2$. The model is completed with hyperpriors: 
$\alpha \sim \Gamma(a_\alpha, b_\alpha)$, $b \sim \text{Unif}(0,1)$, 
${\boldsymbol \mu} \sim \text{N}_2(a_\mu, B_\mu)$, 
${\boldsymbol \Sigma} \sim  \text{IWish}(a_\Sigma, B_\Sigma)$, 
$\lambda \sim \text{N}(a_\lambda, b^2_\lambda)$, $\tau^2 \sim \Gamma^{-1}(a_\tau, b_\tau)$, and 
$\rho \sim \Gamma(a_\rho, b_\rho)$. The hyperparameter values are as follows: 
$a_\alpha = 2$, $b_\alpha = 0.1$, $a_\mu =(3.0, -2.8)'$, $B_\mu= B_\Sigma=((0.19,0)',(0,0.19)')$, 
$a_\Sigma = 4$, $a_\lambda = 0$, $b_\lambda = 42.2$, $a_\tau =a_\rho=a= 2$, $b_\tau = 42.2$, 
and $b_\rho = 2/42.2$.

%\begin{comment}
\begin{figure}[t!]
 \centering
    \includegraphics[width=0.45\textwidth]{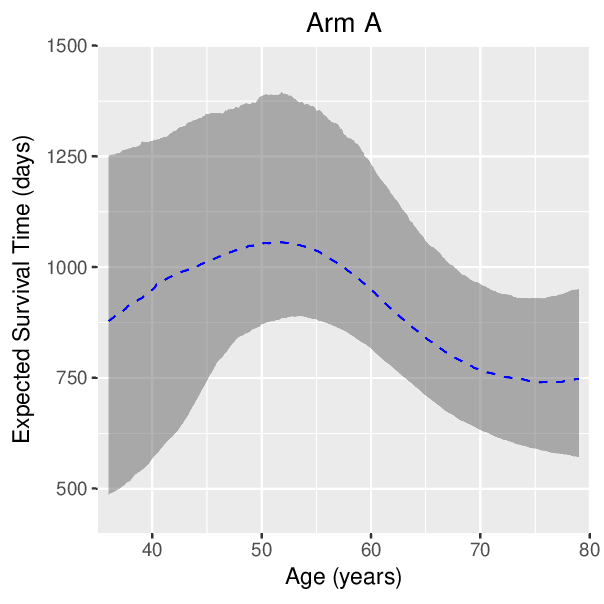}    
    \includegraphics[width=0.45\textwidth]{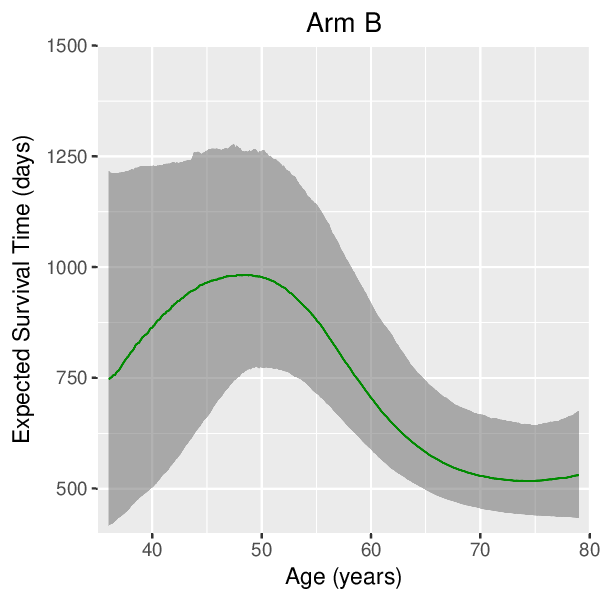}    
\caption{Small cell lung cancer data. Posterior point and $80\%$ interval estimates for
the conditional mean of the survival distribution given patient age at entry (in years), 
for Arm A (left panel) and Arm B (right panel).}
\label{fig:condmean}
\end{figure}
%\end{comment}

\begin{figure}[t!]
 \centering
    \includegraphics[width=0.3\textwidth]{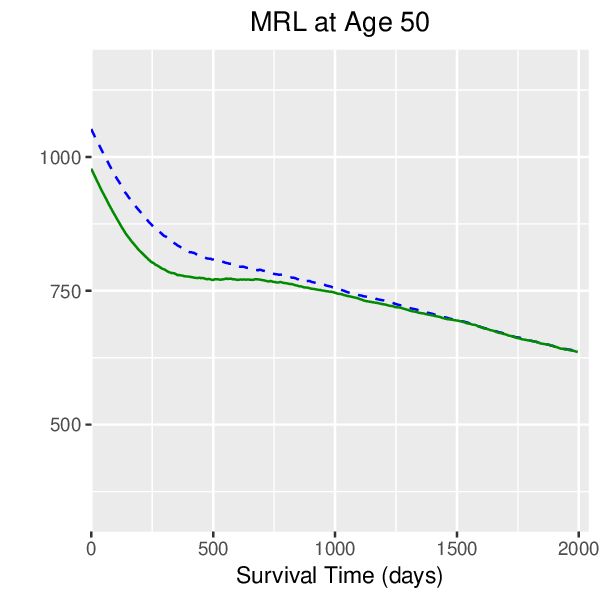}   
    \includegraphics[width=0.3\textwidth]{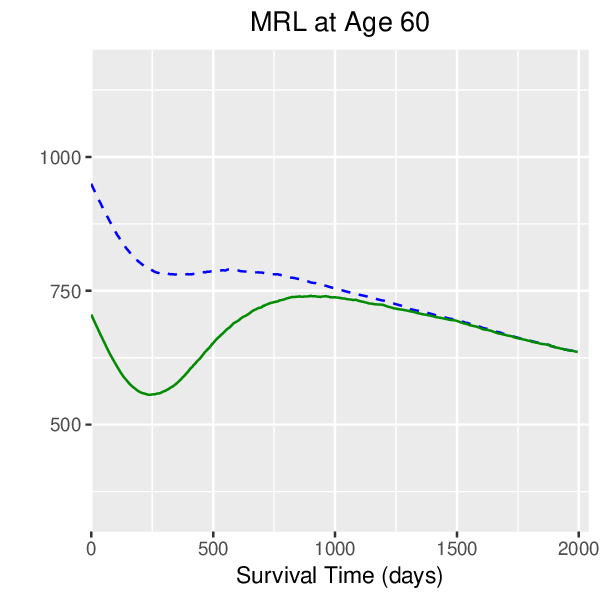}  
    \includegraphics[width=0.3\textwidth]{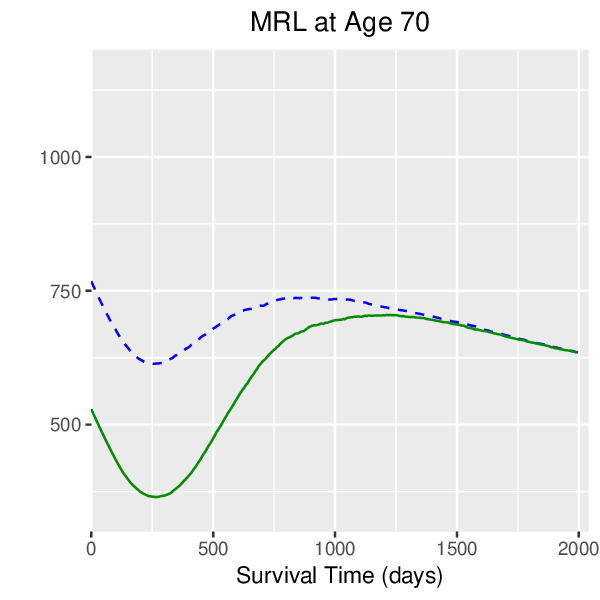} \includegraphics[width=0.3\textwidth]{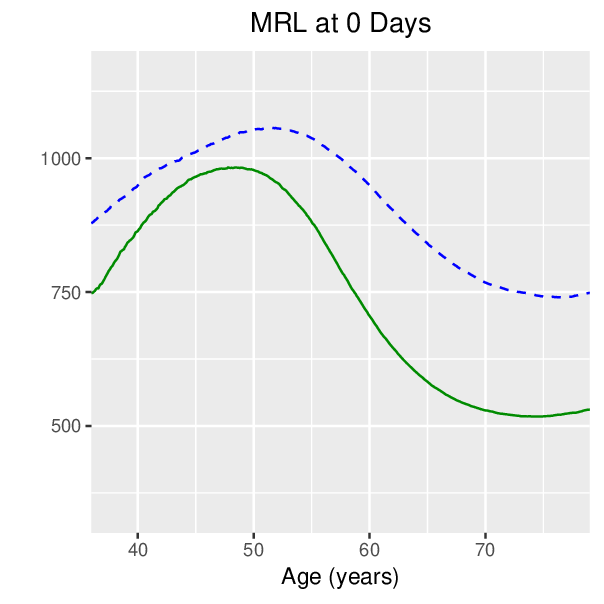} \includegraphics[width=0.3\textwidth]{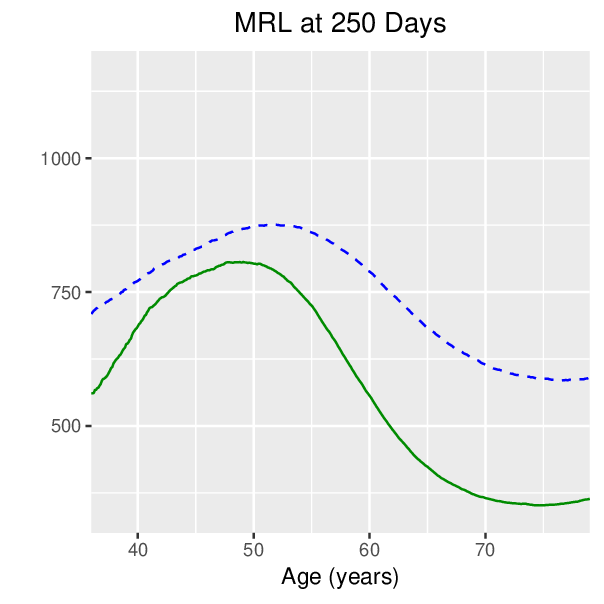} \includegraphics[width=0.3\textwidth]{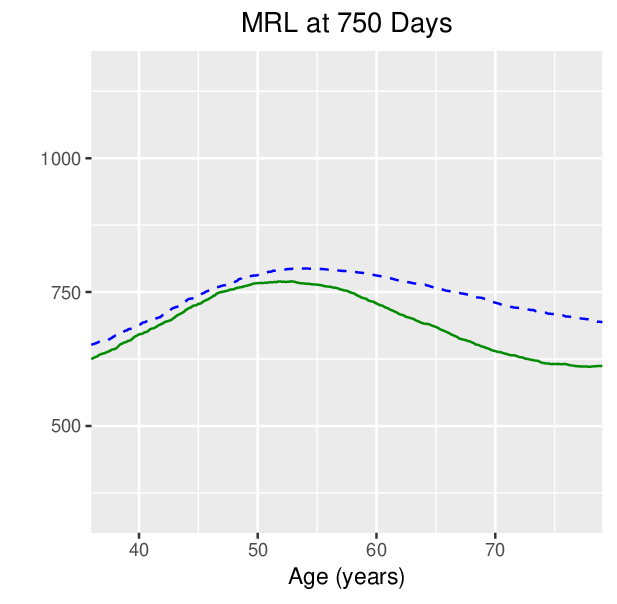} 
\caption{Small cell lung cancer data. Posterior point estimates of the MRL function of Arm A 
(blue dashed lines) and Arm B (green solid lines) for fixed values of age (top panels) and for 
fixed time points (bottom panels).}
\label{fig:MRLreg}
\end{figure}

To our knowledge, earlier work that has considered age as a covariate has done so using linear 
regression. The age-dependent and treatment-dependent weights in the weighted mixture structure 
for the mean regression, $\text{E}(t \mid x, G_s)$, allow for less standard regression relationships 
to be uncovered. Indeed, as shown in Figure \ref{fig:condmean}, the model estimates a non-monotonic 
relationship with age, with an initial increase up to about $50$ years followed by a steeper 
decline, particularly for Arm B. The point estimates for the treatment-specific mean regression 
are also plotted in Figure \ref{fig:MRLreg} (bottom left panel), showing that Arm A achieves 
higher mean survival across the entire range of age values.

Figure \ref{fig:MRLreg} demonstrates the model's capacity to uncover general MRL function shapes
that can change across treatment groups at fixed age values and/or at fixed time points. The top 
panels compare point estimates for Arm A and Arm B mean residual life as a function of time at 
age $50$, $60$, and $70$ years.  At age $50$, the MRL functions are decreasing in time with a 
small separation between treatments up to $750$ days, at which point they become indistinguishable. 
The Arm B estimate suggests a minor dip at about $300$ days. At age $60$, the separation across 
treatments becomes more apparent in the earlier range of survival times, and the dips are more 
pronounced, especially for Arm B; here, the estimates become indistinguishable after about 
$1000$ days. At age $70$, we observe a similar curvature with the age $60$ MRL functions with 
a dip around $250$ days, larger separation for smaller survival times, and essentially identical 
functions after about $1250$ days. While the shape of the MRL functions changes in a non-trivial 
fashion across values of age, Arm A mean residual life remains as high or higher than Arm B.
These results provide a picture about the treatment-specific MRL functions, localized in terms 
of patient's age at entry in the study. In particular, it is interesting to compare the estimates
in the top panels of Figure \ref{fig:MRLreg} with the ones in the bottom right panel of 
Figure \ref{fig:DDPMMabMRL}, based on the model without the age covariate.

Finally, the bottom panels of Figure \ref{fig:MRLreg} compare mean residual life as a function 
of age at three time points: $0$, $250$, and $750$ days. In all three cases, the MRL function
is higher for Arm A across the range of age values. Again, the first case corresponds to mean 
survival as a function of age. Comparing the estimates at $0$ and $250$ days, we observe a similar 
non-monotonic relationship with age, including a similar separation between treatments, although 
there is a decrease in mean residual life at $250$ days. The separation between treatment groups 
and the extent of the non-monotonic trend become less pronounced at $750$ days. Overall, the 
results reinforce the point that Arm A offers a better treatment than Arm B.

\section{Summary}
\label{sec:conc}

We have proposed a nonparametric mixture modeling approach for mean residual life (MRL) regression, 
a problem that, to our knowledge, has not received attention in the Bayesian nonparametrics literature. 
The focus has been on developing general inference methodology for MRL functions across different 
values in the covariate space, as well as for MRL regression relationships across different time points. 
The modeling approach builds from Dirichlet process mixture density regression, including dependent 
Dirichlet process priors to accommodate data from different experimental groups. The methodology 
has been illustrated with both synthetic and real data examples.

\section*{Appendix}

Appendix A includes simulated data examples for the Dirichlet process mixture model developed 
in Section \ref{sec:curve_reg}. Appendix B details the structure and correlation properties 
of the dependent Dirichlet process prior model of Section \ref{subsec:ddpform}, the MCMC 
posterior simulation method, and the derivation of the CPO values used in the model comparison 
of Section \ref{sec:cpo}. 
%These Web Appendices are available along with this paper at the Biometrics website on 
%Wiley Online Library.\vspace*{-8pt}

%  Here, we create the bibliographic entries manually, following the
%  journal style.  If you use this method or use natbib, PLEASE PAY
%  CAREFUL ATTENTION TO THE BIBLIOGRAPHIC STYLE IN A RECENT ISSUE OF
%  THE JOURNAL AND FOLLOW IT!  Failure to follow stylistic conventions
%  just lengthens the time spend copyediting your paper and hence its
%  position in the publication queue should it be accepted.

%  We greatly prefer that you incorporate the references for your
%  article into the body of the article as we have done here 
%  (you can use natbib or not as you choose) than use BiBTeX,
%  so that your article is self-contained in one file.
%  If you do use BiBTeX, please use the .bst file that comes with 
%  the distribution.

\section*{A. Simulation examples for the density regression model}

We provide two simulation examples to demonstrate the capacity of the density regression model
(Section 2 of the main paper) to capture a variety of MRL functional shapes. 

Both examples involve a single continuous covariate. For the first example, we work with a 
finite mixture for the joint response-covariate distribution, specified such that the MRL function takes on various non-standard shapes at different parts of the covariate space. In the second example, we consider an exponentiated Weibull distribution \citep{mudholkar-strivasta:1993} for the survival responses. This is a three-parameter extension of the Weibull distribution that achieves more general shapes for the hazard rate and MRL function. The regression model for the simulation truth is built by defining the three response distribution parameters through specific functions of covariate values, which are drawn from a uniform distribution. The two simulation scenarios are designed to correspond to a setting similar to the model structure, as well as a much more structured parametric setting for the data generating stochastic mechanism. We work with relatively large sample sizes ($1500$ and $500$ for the first and second example) so that the data sets provide reasonably accurate representations of the simulation truth, thus rendering comparison with true MRL functions meaningful. The synthetic data examples of Section 3.2 and the analysis of the real data in Section 4 of the main paper illustrate model inferences under smaller sample sizes. 

We apply the same DP mixture model to both synthetic data sets, with a gamma and normal product kernel, $k(t, x \mid \eta,\phi,\beta,\kappa^2) \propto$ $t^{e^\eta -1}\text{exp}(-e^\phi t)\text{exp}(-0.5\kappa^{-2}(x - \beta)^2)$. The DP centering distribution is defined by $G_{0}(\eta,\phi,\beta,\kappa^2)=$ $\text{N}_{2}((\eta,\phi) \mid {\boldsymbol \mu}, {\boldsymbol \Sigma}) \, \text{N}(\beta \mid\lambda, \tau^2) \, \Gamma^{-1}(\kappa^2 \mid a, \rho)$, where $\Gamma^{-1}(c,d)$ denotes the inverse-gamma distribution with mean $d/(c-1)$ (provided $c>1$). The model is completed with the following hyperpriors: ${\boldsymbol \mu} \sim \text{N}_2(a_\mu, B_\mu)$, ${\boldsymbol \Sigma} \sim \text{IWish}(a_\Sigma, B_\Sigma)$, $\lambda \sim  \text{N}(a_{\lambda}, b_{\lambda})$, $\tau^2 \sim \Gamma^{-1}(a_\tau, b_\tau)$, $\rho \sim \Gamma(a_\rho, b_\rho)$, and $\alpha \sim \Gamma(\alpha \mid a_\alpha, b_\alpha)$, where $\Gamma(c,d)$ denotes the gamma distribution with mean $c/d$. For both examples, we set $a_\alpha = 3$, $b_\alpha = 0.1$, and $L=80$ for the DP truncation level.

\vspace{3pt}
%\subsection{Simulation 1}
\noindent{\bf Simulation 1}

\begin{figure}[t!]
 \centering
    \includegraphics[width=0.45\textwidth]{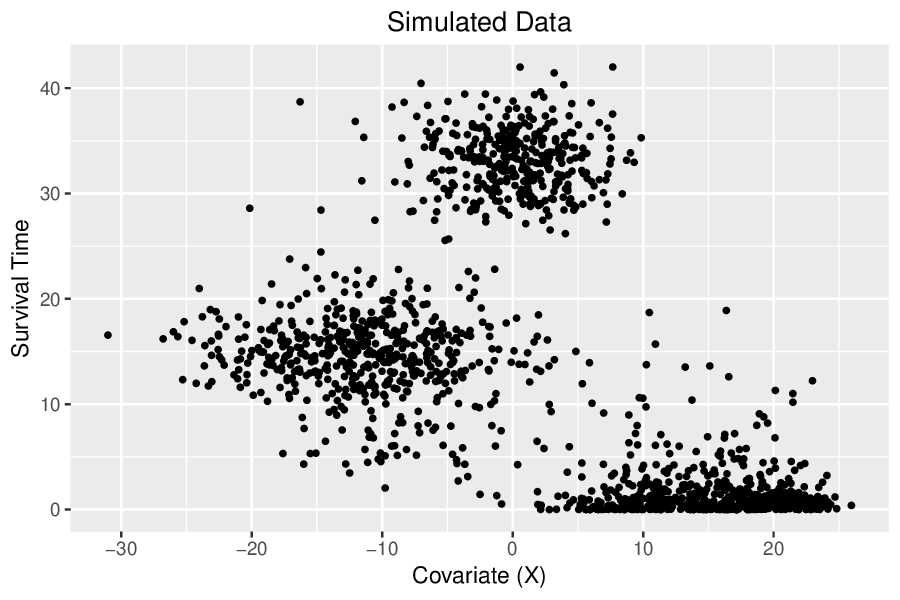}
          \includegraphics[width=0.45\textwidth]{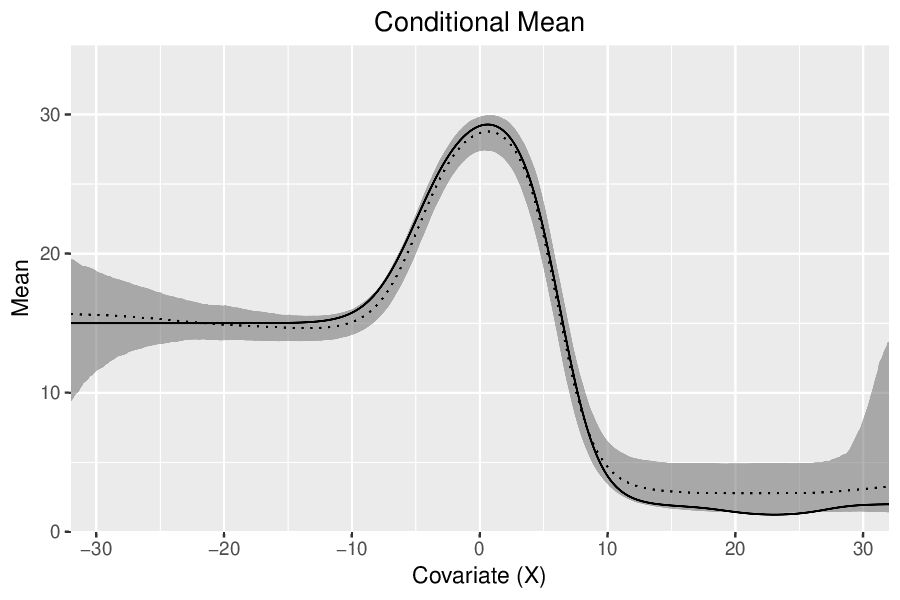}
\caption{Simulated data from the finite mixture. The left panel plots
the data. The right panel shows point (dotted line) and interval
estimates (gray bands) of $\text{E}(T \mid x,G)$, overlaid on the true 
conditional expectation (solid line).}
\label{fig:curvefitsim}
\end{figure}

We simulate $1500$ observations from a population with density: $f(t,x)=$ $\sum_{l=1}^{6} q_l\Gamma(t\mid a_l, b_l) \text{N}(x \mid m_l,s^2_l)$, where $\{ a_{l} \}=$ $(45, 3, 125, 0.4,0.5, 4)$, $\{ b_{l} \}=$ $(3,0.2, 3.8, 0.2,0.3,5)$, $\{ m_{l} \}=$ $(-12, -8, 0 ,12, 18, 21)$, $\{ s_{l} \}=$ $(6, 5, 4, 5, 3, 2)$, and $\{ q_{l} \}=$ $(0.28, 0.1, 0.25, 0.21, 0.11, 0.05)$. The simulated data is shown in the left panel of Figure \ref{fig:curvefitsim}.  The following hyper priors were assumed: $a_\mu= (0.59, -2.12)$, $B_\mu=B_\Sigma = ((0.019,0)', (0, 0.019)')$, $a_\lambda=0$, $a_\tau = 2$, $a_\rho=1$, $b_\lambda=b_\tau=88$, $b_\rho=1/88$.

\begin{figure}[t!]
 \centering
        \includegraphics[width=0.45\textwidth]{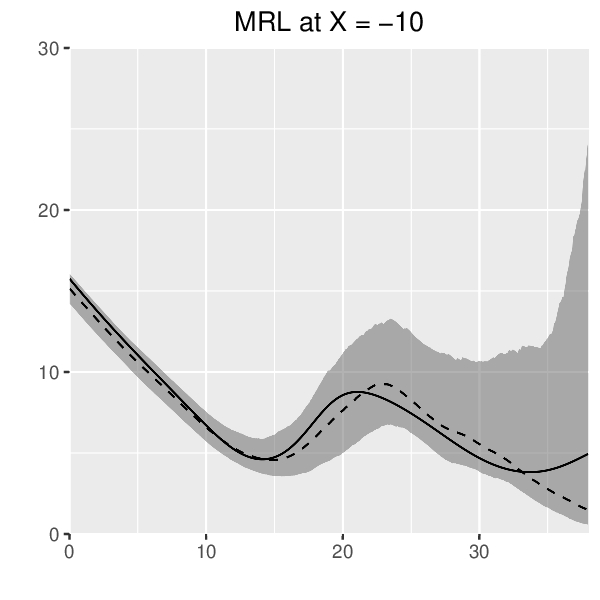} 
        \includegraphics[width=0.45\textwidth]{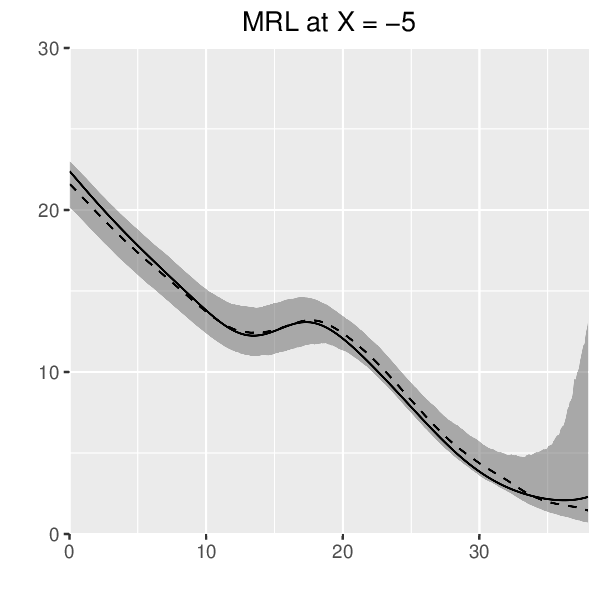}
        \includegraphics[width=0.45\textwidth]{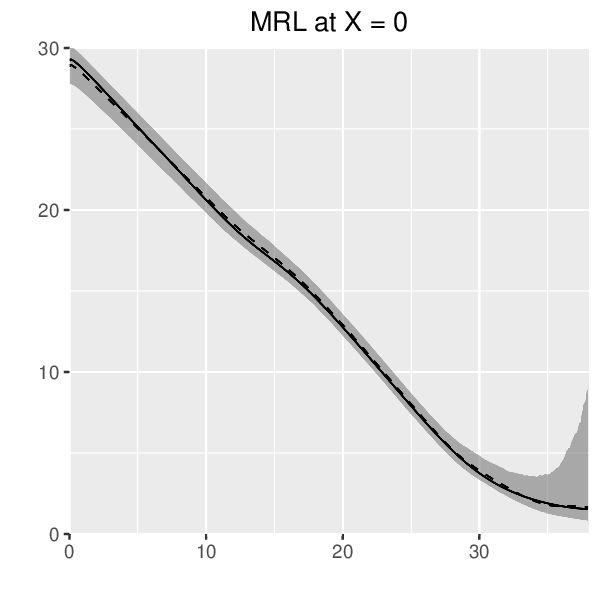}
        \includegraphics[width=0.45\textwidth]{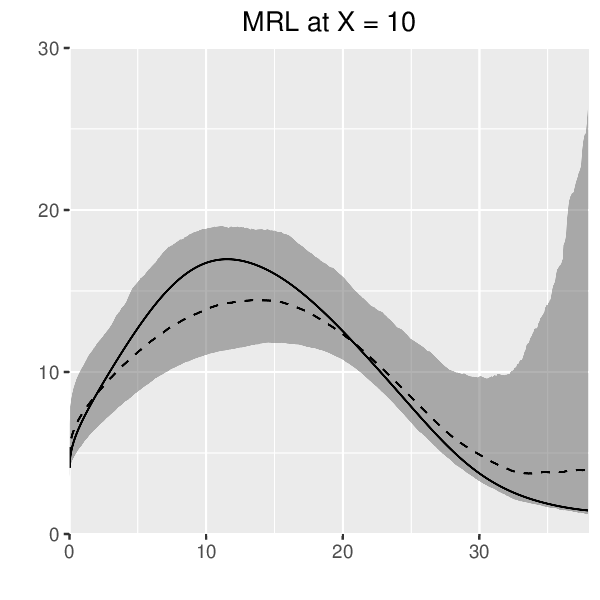}
           \caption{Simulated data from the finite mixture.
Point (dashed line) and $95\%$ interval estimates (gray bands) of the MRL 
function for the specified covariate value overlaying the true MRL
function of the population (solid line).}\label{fig:curvefitmrl}
\end{figure}

The mean of the survival times across a grid of covariate values is shown in Figure~\ref{fig:curvefitsim} (right panel).  In general, the model is able to capture the non-linear trend of the mean over the covariate values.  The truth is captured within the $95\%$ interval estimate save for a small sliver barely outside the interval near the right tail of the covariate space where data is sparse.  The results for MRL functional inference is shown in Figure~\ref{fig:curvefitmrl}.  We provide point and $95\%$ interval estimates for the MRL function at four different covariate values.  The model is able to capture the overall shape of the true MRL functions, despite the variety of and often complexity of the shapes.  At covariate values where the data is most dense, such as $x=-5$ and $x=0$, the inference is more precise as is seen in the narrow interval bands.  As we move to covariate values where data is more sparse, the wide interval bands reflect the uncertainty of the MRL functional shape.       

\vspace{3pt}
%\subsection{Simulation 2}
\noindent{\bf Simulation 2}

\begin{figure}[t!]
 \centering
    \includegraphics[width=0.45\textwidth]{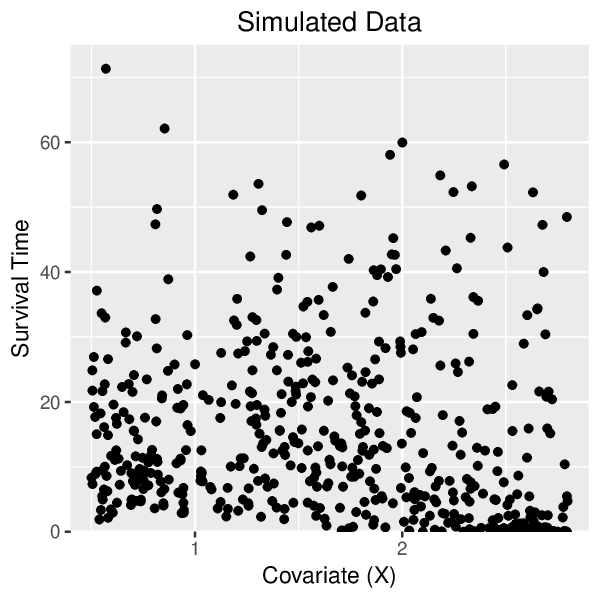}
    \includegraphics[width=0.45\textwidth]{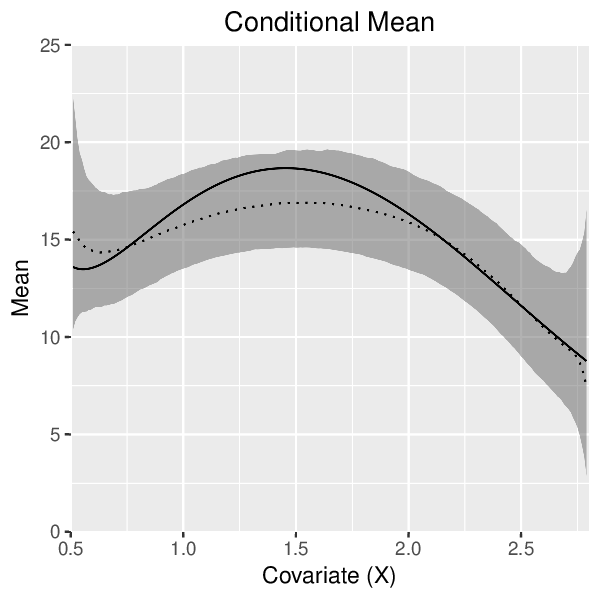}
\caption{Simulated data from the exponentiated Weibull regression model. The left panel plots the data. The right panel shows point (dotted line) and interval estimates (gray bands) of $\text{E}(T \mid x,G)$, overlaid on the true conditional expectation (solid line).}
\label{fig:ExpWeibsim}
\end{figure}

The exponentiated Weibull population has survival function, $S(t \mid \alpha', \theta', \sigma' )=$ $1- [1- \text{exp}\{ - (t/\sigma')^{\alpha'}\}]^{\theta'}$.  The MRL function associated with this distribution can take on increasing, decreasing, constant, upside-down bathtub, and bathtub shapes depending on the shape parameters, $\alpha'$ and $\theta'$, as well as their product ($\sigma'$ is a scale parameter).  We sample $500$ observations from an exponentiated Weibull population with $\alpha' = X$, $\theta' =\text{exp}(2.93 -1.96X)$, and $\sigma' = 14\text{log}(X^3 + 1)$, where $X \sim \text{Unif}(0.5, 2.8)$.  The simulated data is shown in the left panel of Figure \ref{fig:ExpWeibsim}.  The following hyper priors were assumed: $a_\mu= (2.0, -0.8)$, $B_\mu=B_\sigma = ((0.11,0)', (0, 0.11)')$, $a_\lambda=0$, $a_\tau = 2$, $a_\rho=1$, $b_\lambda=b_\tau=4.6$, $b_\rho=1/4.6$.

\begin{figure}[t!]
 \centering
        \includegraphics[width=0.45\textwidth]{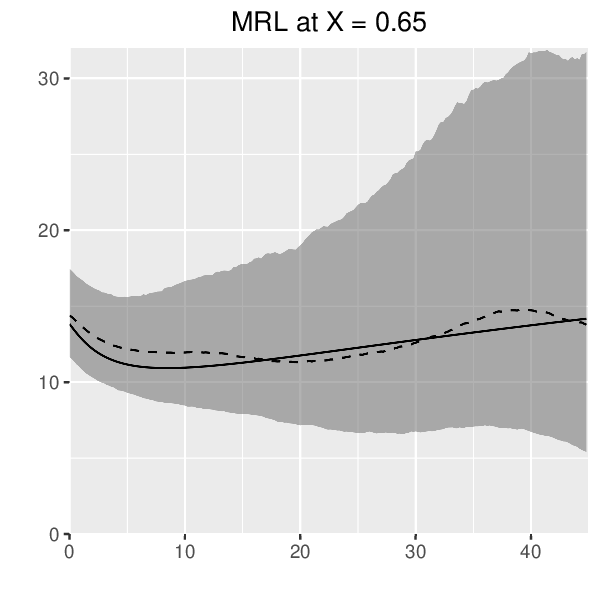} 
           \includegraphics[width=0.45\textwidth]{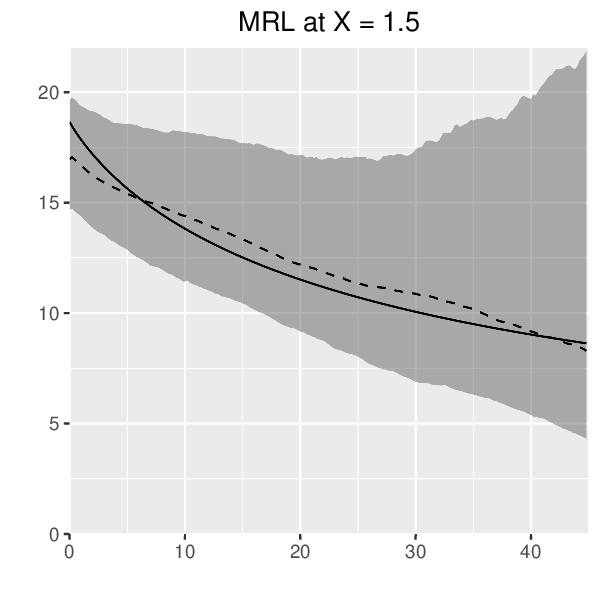}
        \includegraphics[width=0.45\textwidth]{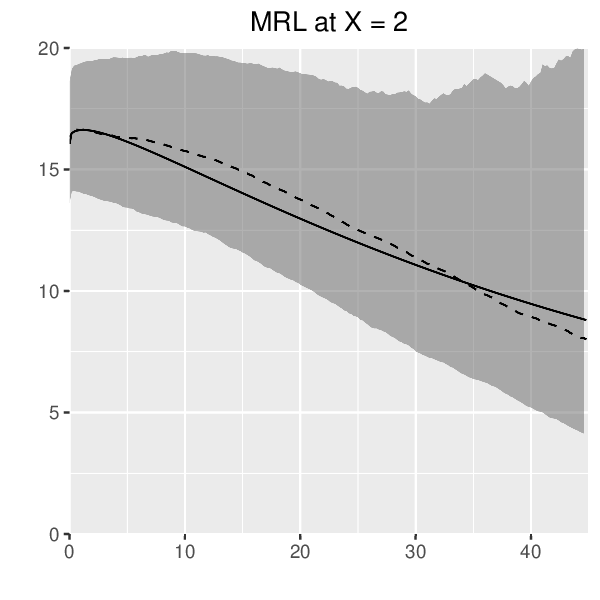}
        \includegraphics[width=0.45\textwidth]{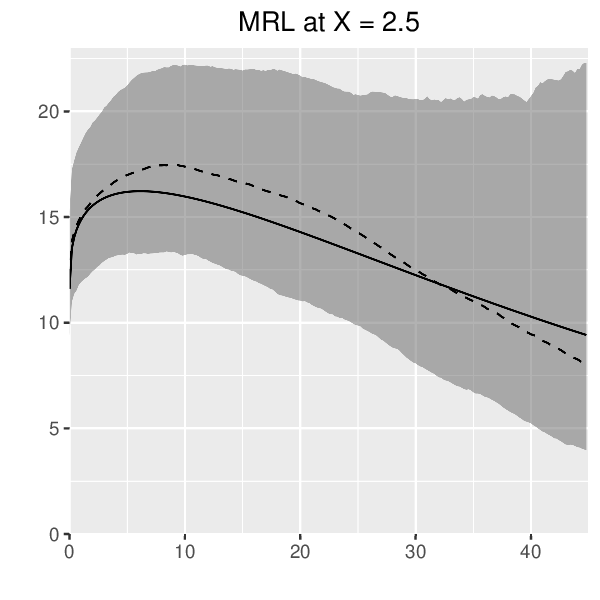}
\caption{Simulated data from the exponentiated Weibull regression model. Point (dotted line) and $95\%$ interval estimates (gray bands) of the MRL function for the specified covariate value overlaying the true MRL function of the population (solid line).}\label{fig:ExpWeibmrl}
\end{figure}

The mean of the survival times across a grid of covariate values is shown in Figure~\ref{fig:ExpWeibsim} (right panel).  Once again, the true mean regression exhibits a non-linear trend that is increasing until about $x = 1.5$ then decreases.  The is captured well within the $95\%$ interval estimate and the parabolic shape is clearly mimicked by the point estimate. The results for MRL functional inference is shown in Figure~\ref{fig:ExpWeibmrl} at four covariate values.  In all four scenarios, the truth is captured within the $95\%$ interval bands while the general shapes are mimicked by the point estimates.

\vspace{6pt}

\section*{B. DDP mixture model}

\subsection*{B.1 Properties of the DDP mixture model}
%\noindent{\Large \textbf{B.1 Properties of the DDP mixture model}}

Here we study the correlation structure under the dependent Dirichlet process (DDP) mixture model (Section 3.1 of the main paper) that is induced by the bivariate beta distribution in \citet{nadarajah-kotz:2005}.  This bivariate beta distribution is based off of the product of independent beta distributions: $U_l\mid \alpha, b \sim \text{Beta}(\alpha, 1-b)$, $V_l \mid \alpha, b \sim \text{Beta}(\alpha, 1-b)$, $W_l \mid \alpha, b \sim \text{Beta}(\alpha +1 -b, b)$.  Letting $\zeta_{lC} = U_lW_l$ and $\zeta_{lT} = V_lW_l$ provides the bivariate beta distribution on $\{\zeta_{lC},\zeta_{lT}\}$. Generating $\{\zeta_{lC},\zeta_{lT}\}$ for $l \in \{1,2, ...\}$ under independent and identical processes, the component weights of the DDP mixture model are then defined for group $s =\{C, T\}$ by $w_{1s}=1-\zeta_{1s}$, $w_{ls}=(1-\zeta_{ls})\prod_{r=1}^{l-1}\zeta_{rs}$, for $l\in\{2,3,...\}$. 

We are interested in obtaining the correlation between the two mixing distributions, $G_C$ and $G_T$, implied under this bivariate beta distribution.  We first begin with examining the correlation between $\zeta_{C}$ and $\zeta_{T}$, $\text{Corr}(\zeta_{C},\zeta_{T}\mid \alpha, b)$.  We omit the component subscript in the latent variables, since results are the same for each $l\in \{1,2,...\}$. The covariance can be written as, $\text{Cov}(\zeta_{C},\zeta_{T}\mid \alpha, b) = \text{E}(\zeta_{C}\zeta_{T}\mid \alpha, b) - \text{E}(\zeta_{C}\mid \alpha, b)\text{E}(\zeta_{T}\mid \alpha, b) = \text{E}((UW\mid \alpha, b)(VW\mid \alpha, b)) - \text{E}(UW\mid \alpha, b)E(VW\mid \alpha, b)$.  Using the fact that $U,V,W$ are independent, the covariance becomes $\text{E}(U\mid \alpha, b)\text{E}(V\mid \alpha, b)\text{E}(W^2\mid \alpha, b) - \text{E}(U\mid \alpha, b)\text{E}(V\mid \alpha, b)\text{E}^2(W\mid \alpha, b) = \text{E}(U\mid \alpha, b)\text{E}(V\mid \alpha, b)\text{Var}(W\mid \alpha, b)$. Now, since $\zeta_C$ and $\zeta_T$ have the same marginal distribution, $\text{Beta}(\alpha,1)$,  the covariance and correlation reduces to:
\begin{eqnarray*} \text{Cov}(\zeta_{C},\zeta_{T}\mid \alpha, b)&=&  \frac{\alpha^2b}{(\alpha +1 - b)(\alpha +1)^2(\alpha +2)} \nonumber \\
  \text{Corr}(\zeta_{C},\zeta_{T}\mid \alpha, b)&=& \frac{\alpha b}{\alpha+1-b} 
\end{eqnarray*}

\noindent The correlation between $\zeta_C$ and $\zeta_T$ can take values on the interval $(0,1)$.  As $b\to0$ and/or $\alpha\to 0$, the correlation goes to $0$.  As $b\to 1$ and/or $\alpha\to \infty$, the correlation tends to $1$. 

We next explore the correlation of the weights, $\text{Corr}(w_{lC},w_{lT}\mid \alpha, b)$ for $l\in\{1,2,....\}$.  When $l=1$, $w_{1s} = 1-\zeta_{1s}$, which is simply a linear operation, hence the covariance and correlation are the same as before.  The $\text{Cov}(w_{1C},w_{1T}\mid \alpha, b) = \text{Cov}(\zeta_{C},\zeta_{T}\mid \alpha, b)$ and $\text{Corr}(w_{1C},w_{1T}\mid \alpha, b) = \text{Corr}(\zeta_{C},\zeta_{T}\mid \alpha, b)$ are given above.  The case is different for $l=\{2,3,...\}$.  In this case, the covariance is defined as $\text{E}(((1-\zeta_{lC})\prod_{r=1}^{l-1}\zeta_{rC}\mid)((1-\zeta_{lT})\prod_{r=1}^{l-1}\zeta_{rT})\mid \alpha, b) - \text{E}( (1-\zeta_{lC})\prod_{r=1}^{l-1}\zeta_{rC}\mid \alpha, b)\text{E}((1-\zeta_{lT})\prod_{r=1}^{l-1}\zeta_{rT}\mid \alpha, b)$.  Using the fact that $\zeta_{ls}$ are independent across $l=1,...,L$, for each $s\in\{C,T\}$, the covariance, for $l\in\{2,3,...\}$, can be expressed as
\begin{eqnarray*}
 \text{Cov}(w_{lC},w_{lT}\mid \alpha, b)&=&\frac{(\alpha +1 -b)(\alpha+2)+\alpha^2b}{(\alpha +1 -b)(\alpha+1)^2(\alpha+2)}\left(\frac{\alpha^2b + \alpha^2(\alpha +1 -b)(\alpha+2)}{(\alpha +1 -b)(\alpha+1)^2(\alpha+2)}\right)^{l-1} \nonumber \\
 && - \frac{1}{(\alpha+1)^2}\left(\frac{\alpha^2}{(\alpha+1)^2}\right)^{l-1}
\end{eqnarray*} 
The variance for the weights are independent of group, and can be expressed as $\text{Var}(w_{ls}\mid \alpha, b) = 2/(\alpha+1)(\alpha+2)[(\alpha+\alpha^2(\alpha+2))/((\alpha+1)^2(\alpha+2))]^{l-1} -1/(\alpha+1)^2[\alpha^2/(\alpha+1)^2]^{l-1}$. Therefore, the correlation, for $l\in\{2,3,...\}$, can be obtained by $\text{Corr}(w_{lC},w_{lT}\mid \alpha, b) = \text{Cov}(w_{lC},w_{lT}\mid \alpha, b)/\text{Var}(w_{ls}\mid \alpha, b)$, which is in closed form, but does not reduce. The correlation between the weights for $l\in\{2,3,...\}$ also takes values on the interval $(0,1)$ and behaves the same in terms of the limits of $\alpha$ and $b$ as in the case when $l=1$. The component value, $l$, plays a slight role in the correlation, specifically as $l$ get larger, the rate of change for smaller $\alpha$ values becomes less extreme.        

%%%%%%%GcGT

We now turn to the correlation between the two mixing distributions, $G_C$ and $G_T$.  
Let $B$ represent a specified (measurable) set in the joint space of the mixing parameters.  
%In the model we present in Section 3.1 of the main paper, ${\boldsymbol \Theta}$ is equivalent 
%to $\mathbb{R}^2$, so $B$ is simply a subset of $\mathbb{R}^2$.}
%\footnote{
%\textcolor{blue}{This is not quite right, we also have the parameters for the kerneldensity associated with covariates; either revise or just remove the sentence in blue}}
Recall that the mixing distribution for group $s$ has form $G_s(B)= \sum_{l=1}^\infty w_{ls}\delta_{{\boldsymbol \theta}_l}(B)$.  Marginally, $G_s(B)$ follows a DP, so the expectation and variance of $G_s(B)$ is $G_0(B)$ and $G_0(B)[1-G_0(B)]/(\alpha+1)$, respectively.  The covariance between $G_C(B)$ and $G_T(B)$ is given by $Cov\left(\sum_{l=1}^\infty w_{lC} \delta_{{\boldsymbol \theta}_l}(B),\right.$ $\left.\sum_{l=1}^\infty w_{lT} \right.$ $\left.\delta_{{\boldsymbol \theta}_l}(B)\mid \alpha, b, G_0(B)\right)$, simplifying to, $G_0(B)\sum_{l=1}^\infty \text{E}(w_{lC}w_{lT}\mid \alpha, b) + 2G_0^2(B)\sum_{l=1}^\infty $ $ \sum_{m=l+1}^\infty \text{E}(w_{lC} w_{mT}\mid \alpha, b) - G_0^2(B)$.
%\footnote{
%\textcolor{blue}{Something is not right here, the expression can't depend on the randomvariables $w_{ls}$, we need the respective covariances for the weights, right? please double check the derivation that leads to Eq. (1) (the result for the correlation is much simpler than what we typically get from DDP priors)}
%}
The infinite series converges under geometric series, and the covariance simplifies to be: 
\begin{eqnarray*}
\text{Cov}(G_C(B),G_T(B)\mid \alpha, b, G_0(B)) &=& G_0(B)(1-G_0(B))\left(\frac{(\alpha-2)b+\alpha +2}{\alpha(2\alpha -3b+5) -2b+2} \right)\end{eqnarray*}
\noindent The correlation does not depend on the choice of $B$ or $G_0$; it is driven by $\alpha$ and $b$ alone:
\begin{eqnarray}
\text{Corr}(G_C(B),G_T(B)\mid \alpha, b) &=& \frac{(\alpha+1)((\alpha-2)b+\alpha +2)}{\alpha(2\alpha -3b+5) -2b+2} \label{eqn:cor}
%=\frac{\alpha^2(b+1)+\alpha(3-b)-2b+2}{2\alpha^2 - 3\alpha b +5\alpha -2b+2}
\end{eqnarray}

The correlation of the mixing distribution lives on the interval $(1/2, 1)$.  As $\alpha \to 0$ and/or $b\to 1$, the correlation tends to $1$.  When $\alpha \to \infty$ the correlation tends to $(b+1)/2$ and as $b \to 0$ the correlation tends to $(\alpha+1)/(2\alpha+1)$, so when $\alpha \to \infty$ and  $b \to 0$ the correlation goes to $1/2$. Although this correlation space is limited, it is a typical range seen in the literature (e.g. \cite{mckenzie:1985}).  It can easily be shown that the correlation of the survival distributions between the two groups given $G_C$ and $G_T$ also live on $(1/2, 1)$, which demonstrates the importance of prior knowledge of the relationship between the distributions of the two group survival times.  

%%%%TcTt

While the possible values of correlation of the mixing distributions of the two groups is restricted to $(1/2,1)$, the correlation between the survival times (marginal of the covariate(s), see Web Appendix B.3) across the two groups, $\text{Corr}(T_C, T_T \mid \alpha, b,{\boldsymbol \mu},{\boldsymbol \Sigma})$, takes on values in $(0,1)$. The $\text{Corr}(T_C, T_T \mid \alpha, b,{\boldsymbol \mu},{\boldsymbol \Sigma})$ is found by marginalizing over the mixing distributions, $G_C$ and $G_T$.  Starting with the covariance, $\text{Cov}(T_C,T_T \mid \alpha,b,{\boldsymbol \mu},{\boldsymbol \Sigma})=\text{E}(T_CT_T \mid \alpha,b,{\boldsymbol \mu},{\boldsymbol \Sigma}) - \text{E}(T_C\mid \alpha,b,{\boldsymbol \mu},{\boldsymbol \Sigma})\text{E}(T_T\mid \alpha,b,{\boldsymbol \mu},{\boldsymbol \Sigma}) = \text{E}(\text{E}(T_C\mid G_C, \alpha,b)\text{E}(T_T\mid G_T, \alpha,b)\mid \alpha,b,{\boldsymbol \mu},{\boldsymbol \Sigma}) - \text{E}(\text{E}(T_C\mid G_C,\alpha,b)\mid \alpha,b,{\boldsymbol \mu},{\boldsymbol \Sigma})\text{E}(\text{E}(T_T\mid G_T, \alpha,b)\mid \alpha,b,{\boldsymbol \mu},{\boldsymbol \Sigma})$.  Under DDP mixture model without covariates, we have the gamma kernel with bivariate normal $G_0$. Thus, the covariance is given by the following,
\begin{eqnarray*}
\text{Cov}(T_C,T_T \mid \alpha,b,{\boldsymbol \mu},{\boldsymbol \Sigma}) &=& \left(e^{t_2'{\boldsymbol \mu} +\frac{1}{2}t_2'{\boldsymbol \Sigma} t_2} - e^{2(t_3'{\boldsymbol \mu} +\frac{1}{2}t_3'{\boldsymbol \Sigma} t_3} \right) \left(\frac{(\alpha-2)b +\alpha +2}{\alpha(2\alpha -3b +5) -2b +2}\right)
\end{eqnarray*}
where $t_2 = (2,-2)'$ and $t_3 = (1,-1)'$.  The variance of $T_s$, for both $s\in\{C,T\}$, conditional on $\alpha$ and $b$ is given by, $e^{t_1'{\boldsymbol \mu} +\frac{1}{2}t_1'{\boldsymbol \Sigma} t_1} +e^{t_2'{\boldsymbol \mu} +\frac{1}{2}t_2'{\boldsymbol \Sigma} t_2} - e^{2(t_3'{\boldsymbol \mu} +\frac{1}{2}t_3'{\boldsymbol \Sigma} t_3)}$ where that $t_1=(1,-2)'$.  Hence, the correlation is given by,
\begin{eqnarray*}
\text{Corr}(T_C,T_T\mid \alpha,b,{\boldsymbol \mu},{\boldsymbol \Sigma}) &=& \left[\left(e^{t_2'{\boldsymbol \mu} +\frac{1}{2}t_2'{\boldsymbol \Sigma} t_2} - e^{2(t_3'{\boldsymbol \mu} +\frac{1}{2}t_3'{\boldsymbol \Sigma} t_3)} \right)\left(\frac{(\alpha-2)b +\alpha +2}{\alpha(2\alpha -3b +5) -2b +2}\right)\right]\left. \middle/ \right. \nonumber  \\
&& \left[e^{t_1'{\boldsymbol \mu} +\frac{1}{2}t_1'{\boldsymbol \Sigma} t_1} +e^{t_2'{\boldsymbol \mu} +\frac{1}{2}t_2'{\boldsymbol \Sigma} t_2} - e^{2(t_3'{\boldsymbol \mu} +\frac{1}{2}t_3'{\boldsymbol \Sigma} t_3)}\right]    
\end{eqnarray*}
As the $e^{t_1'{\boldsymbol \mu} +\frac{1}{2}t_1'{\boldsymbol \Sigma} t_1}= E(e^{\eta -2\phi}\mid {\boldsymbol \mu},{\boldsymbol \Sigma}) \to 0$ 
the correlation simplifies to $((\alpha-2)b +\alpha +2)/(\alpha(2\alpha -3b +5) -2b +2)$.  In this case, as $\alpha \to 0$ the correlation tends to $1$ and as $\alpha \to \infty$ the correlation tends to $0$. Also, as $b \to 0$ the correlation tends to $1/(2\alpha +1)$ and as $b\to 1$ the correlation tends to $1/(\alpha +1)$.  These results are scaled down as $\text{E}(e^{\eta -2\phi}\mid {\boldsymbol \mu},{\boldsymbol \Sigma})$, the expectation of the kernel variance, gets larger.

\vspace{6pt}
\subsection*{B.2 MCMC Details}
%\noindent{\Large \textbf{B.2 MCMC Details}} 

\noindent 
Here we show the posterior sampling algorithm used for the DDP mixture model in the presence 
of a single random continuous (real-valued) covariate, as applied in Section 4.2 of the main paper. 
The mixture kernel density comprises a product of the gamma density for the survival 
response and a normal density for the random covariate.
Omitting the model component for the random covariate yields the model applied in 
Sections 3.3 and 4.1 of the main paper. Assuming a single group, i.e., $s$ having only 
a single index value, will yield the algorithm pertaining to the simulation examples in Web Appendix A. We obtain posterior samples using the blocked Gibbs sampler and working with the latent parameters of the bivariate beta distribution. Posterior samples are based on a truncation approximation, $G_{Ls}$, to $G_s$:  $G_{Ls} = \sum_{l=1}^L p_{ls}\delta_{{\boldsymbol \theta}_l}$.  
Specifically, the atoms are defined as ${\boldsymbol \theta}_l =$
$(\eta_l, \phi_l, \beta_l, \kappa^2_l) \stackrel{\text{i.i.d.}}{\sim} G_0$, for $l=1,...,L$ with corresponding weights $p_{1s} = 1 - \zeta_{1s}, \ p_{ls} = (1- \zeta_{ls})\prod_{r=1}^{l-1} \zeta_{rs}$ for  $l\in\{2,3,...,L-1\}$ with $(\zeta_{lC}, \zeta_{lT})| {\boldsymbol \phi} \stackrel{\text{ind}}{\sim} \mbox{Biv-Beta}(\cdot \mid {\boldsymbol \phi})$, and $p_{Ls} = 1- \sum_{l=1}^{L-1}p_{ls}$.

Upon introducing the latent configuration variables, $\mathbf{w} = \{\mathrm{w}_{is} : i=1,...,n_s \mid s=C,T \}$, such that $\mathrm{w}_{is} = l$ if the $i^{th}$ observation in group $s$ is assigned to mixture component $l$, the full hierarchical version of the model is written as,
\begin{eqnarray*}
(t_{is}, x_{is}) \mid {\mathrm w}_{is}, {\boldsymbol \theta}_l 
&\stackrel{\text{ind}}{\sim}& \Gamma(t_{is}\mid e^{\eta_{{\mathrm w}_{is}}}, 
e^{\phi_{{\mathrm w}_{is}}}) \, \text{N}(x_{is}\mid \beta_{{\mathrm w}_{is}}, 
\kappa^2_{{\mathrm w}_{is}}) \\
{\mathrm w}_{is} \mid \{(\zeta_{ls})\} &\stackrel{\text{ind}}{\sim} & 
\sum_{l=1}^L\{(1- \zeta_{ls}) \prod\nolimits_{r=1}^{l-1}\zeta_{rs}\}\delta_l({\mathrm w}_{is}), 
\ \ \ \mbox{for} \  i=1,...,n_s \  \mbox{and} \ s\in\{C,T\} \\
\{(\zeta_{lC},\zeta_{lT})\}\mid \alpha, b & 
\sim& \text{Biv-Beta}(\{(\zeta_{lC},\zeta_{lT})\}\mid \alpha,b)\\
({\eta}_l, \phi_l)' \mid {\boldsymbol \mu},{\boldsymbol \Sigma} & \stackrel{\text{i.i.d.}}{\sim} & 
\text{N}_2(({\eta}_l, \phi_l)' \mid {\boldsymbol \mu},{\boldsymbol \Sigma}), 
\ \ \mbox{for} \ l=1,...,L  \\
(\beta_l, \kappa^2_l) \mid \lambda, \tau^{2}, \rho & \stackrel{\text{i.i.d.}}{\sim} & 
\text{N}(\beta_l \mid \lambda, \tau^2) \, \Gamma^{-1}(\kappa^{2}_{l} \mid a, \rho),
\ \ \mbox{for} \ l=1,...,L 
\end{eqnarray*}
where $\zeta_{lC} = U_{l} \, W_{l}$ and $\zeta_{lT} = V_{l} \, W_{l}$, for $l\in\{1,...,L\}$,
with $U_{l} \mid \alpha,b \stackrel{\text{i.i.d.}}{\sim} \text{Beta}(\alpha, 1-b)$, 
$V_{l} \mid \alpha,b \stackrel{\text{i.i.d.}}{\sim} \text{Beta}(\alpha, 1-b)$, and  
$W_{l} \mid \alpha,b \stackrel{\text{i.i.d.}}{\sim} \text{Beta}(1+\alpha - b, b)$. 
We place the following hyperpriors: 
$\alpha \sim \Gamma(\alpha \mid a_\alpha, b_\alpha)$, $b \sim \text{Unif}(b\mid 0,1)$, 
${\boldsymbol \mu} \sim \text{N}_2(\boldsymbol{\mu}\mid a_\mu, B_\mu)$, 
${\boldsymbol \Sigma} \sim  \text{IWish}({\boldsymbol \Sigma}\mid a_\Sigma, B_\Sigma)$, 
$\lambda \sim \text{N}(\lambda\mid a_\lambda, b^2_\lambda)$, 
$\tau^2 \sim \Gamma^{-1}(\tau^2\mid a_\tau, b_\tau)$, and 
$\rho \sim \Gamma(\rho\mid a_\rho, b_\rho)$.

%Let $L_s^{*}$  be the number of distinct components, and ${\mathbf w}_{s}^*\equiv \{{\mathrm w}^*_{js}: j=1,..., L^*_s\}$ be the vector of latent configuration variables for group $s\in\{C,T\}$. 
For subject $i =1,...,n_s$, let $\delta_{is} = 0$ if $t_{is}$ is observed and $\delta_{is} = 1$ if $t_{is}$ is right censored. Let $\Psi$ represent the vector of the most recent iteration of all other parameters.  Let $b=1,...,B$ be the number of iterations in the MCMC.  The posterior samples of $p({\boldsymbol \eta}, {\boldsymbol \phi}, {\boldsymbol \beta}, {\boldsymbol \kappa}^2,{\mathbf w},{\boldsymbol \zeta},{\boldsymbol \mu},{\boldsymbol \Sigma}, \lambda, \tau^2, \rho, \alpha, b\mid data)$ can be obtained by the following:

First, we consider updates for $({\eta}_l,\phi_l)'$,$\beta_l$, and $\kappa^2_l$ for $l= 1,..., L $. If  $ l \notin  {\mathbf w}_C^{*(b)} \cup {\mathbf w}_T^{*(b)} $, then draw $p({\eta}^{(b+1)}_l, \phi^{(b+1)}_l\mid {\Psi}, data)  \stackrel{\mbox{}}{\sim} N_2({\boldsymbol \mu}^{(b)},{\boldsymbol \Sigma}^{(b)})$, $p(\beta_l^{(b+1)}\mid data, \Psi) \stackrel{\mbox{}}{\sim} N(\lambda^{(b)}, \kappa_l^{2(b)})$, and  $p(\kappa_l^{2(b+1)} \mid {\Psi}, data) \stackrel{\mbox{}}{\sim} \Gamma^{-1}(a, \rho^{(b)})$. If $ l \in  {\mathbf w}_C^{*(b)} \cup l \in  {\mathbf w}_T^{*(b)}$,  we have $p(\eta_l, \phi_l\mid {\Psi}, data) \propto$ $ N_2((\eta_l, \phi_l)' \mid {\boldsymbol \mu}, {\boldsymbol \Sigma})\prod_{s\in\{C,T\}}\prod_{\{i:l={\mathrm w}_{is}\}}[\Gamma(t_{is}\mid e^{\eta_l}, e^{\phi_l})]^{1-\delta_{is}}[\int_{t_{is}}^\infty \Gamma(u_i\mid e^{\eta_l}, e^{\phi_l})dt_i]^{\delta_{is}}$.  We use a Metropolis-Hastings step with proposal distribution $(\eta_l', \phi_l')' \sim N_2((\eta_l^{(b)}, \phi_l^{(b)})', cS^2)$, where $S^2$ is updated from the average posterior samples of  ${\boldsymbol \Sigma}$ under initial runs, and $c>1$. 

For $\beta_l$ and $\kappa_l$, we have $p(\beta_l \mid {\Psi}, data) \propto  N(\beta_l \mid \lambda, \tau^2) \prod_{s\in\{C,T\}}\prod_{\{i:l={\mathrm w}_{is}\}}N(x_{is}\mid \beta_l, \kappa^2_l)$ and $p(\kappa_l^2\mid data, \Psi) \propto \Gamma^{-1}(\kappa_l^2\mid a, \rho)  \prod_{s\in\{C,T\}} $ $\prod_{\{i:l={\mathrm w}_{is}\}}N(x_{is}\mid \beta_l, \kappa^2_l)$. Thus, we sample via:
\begin{eqnarray*}
&&p(\beta_l^{(b+1)}\mid {\Psi}, data) \stackrel{\mbox{}}{\sim} N(m_\beta, s^2_\beta)  \\
&&p(\kappa_l^{2(b+1)} \mid {\Psi}, data) \stackrel{\mbox{}}{\sim} \Gamma^{-1} \left(a + 0.5\sum_{s\in\{C,T\}}M_{ls},  \rho^{(b)} + 0.5\sum_{s\in\{C,T\}}\sum_{\{i:l={\mathrm w}_{is}\}} (x_{is} - \beta_l^{(b+1)})^2 \right) 
\end{eqnarray*}
\noindent where $m_\beta = s^2_\beta \left(\kappa_l^{-2(b)} \left[\sum_{s\in\{C,T\}}\sum_{\{i:l={\mathrm w}_{is}\}} x_{is}\right] + \tau^{-2(b)}\lambda^{(b)}\right)$, $s^2_\beta = \left(\tau^{-2(b)} + \kappa_l^{-2(b)}\left[\sum_{s\in\{C,T\}}\right. \right.$ $\left. \left. M_{ls}\right]\right)^{-1}$, and $M_{ls} = \sum_{\{i:l={\mathrm w}_{is}\}} 1$, for $s\in\{C,T\}$.

\vspace{.2in}

To obtain samples from $p({\boldsymbol \zeta}\mid  {\Psi}, data)$ we work with $\{U_l,V_l,W_l\}$.  Using slice sampling, we can introduce latent variables $\nu_l$ and $\gamma_l$ for $l=1,...,L$.  The Gibbs steps are given by:
\begin{eqnarray*}
&&p(\nu_l^{(b+1)}\mid {\Psi},data) \sim \text{Unif}\left(0,(1-U^{(b)}_lW^{(b)}_l)^{M^{(b)}_{lC}} \right)\\
&&p(\gamma_l^{(b+1)}\mid {\Psi},data) \sim \text{Unif}\left(0,(1-V^{(b)}_l W^{(b)}_l)^{M^{(b)}_{lT}}\right) \\
&&p(U_l^{(b+1)}\mid \Psi, data) \sim \text{Beta}\left((\sum_{r=l+1}^L M^{(b)}_{rC}) +\alpha, 1-b\right){\boldsymbol 1}_{\left(0, \frac{1}{W^{(b)}_l}\left[1- exp\left(\frac{log(\nu^{(b+1)}_l)}{M^{(b)}_{lC}}\right)\right]\right)}\\
&&p(V_l^{(b+1)}\mid \Psi, data) \sim \text{Beta}\left((\sum_{r=l+1}^L M^{(b)}_{rT}) +\alpha, 1-b\right){\boldsymbol 1}_{\left(0, \frac{1}{W^{(b)}_l}\left[1- exp\left(\frac{log(\gamma^{(b+1)}_l)}{M^{(b)}_{lT}}\right)\right]\right)}\\
&&p(W_l^{(b+1)}\mid \Psi, data) \sim \text{Beta}\left((\sum_{r=l+1}^L  M^{(b)}_{rT}+ M^{(b)}_{rC}) +\alpha +1 -b, b\right){\boldsymbol 1}_{\left(0,m^*\right)}
\end{eqnarray*}

\noindent where  $m^* = \text{min}\left\{ \frac{1}{U^{(b+1)}_l}\left[1- exp\left(\frac{log(\nu^{(b+1)}_l)}{M^{(b)}_{lC}}\right)\right], \frac{1}{V^{(b+1)}_l}\left[1- exp\left(\frac{log(\gamma^{(b+1)}_l)}{M^{(b)}_{lT}}\right)\right]\right\}\\
\mbox{Set} \ \zeta^{(b+1)}_{lC} = U^{(b+1)}_lW^{(b+1)}_l$ and $\zeta^{(b+1)}_{lT} = V^{(b+1)}_lW^{(b+1)}_l 
$\\

For the update of ${\mathrm w}_{is}$ we have $p({\mathrm w}_{is}\mid {\Psi}, data) \propto  \Gamma(t_{is}\mid e^{\eta_{{\mathrm w}_{is}}}, e^{\phi_{{\mathrm w}_{is}}})\text{N}(x_{is}\mid \beta_{{\mathrm w}_{is}}, \kappa^2_{{\mathrm w}_{is}})  \sum_{l=1}^L\{(1 $ $- \zeta_{ls}) \prod_{r=1}^{l-1}\zeta_{rs}\}\delta_l({\mathrm w}_{is})$, so we sample from  $p({\mathrm w}^{(b+1)}_{is}\mid {\Psi}, data)  \stackrel{}{\sim}  \sum_{l=1}^L\tilde{p}_{lis}\delta_{(l)}({\mathrm w}_{is})$ where $\tilde{p}_{lis} = p_{ls}[\Gamma(t_{is}\mid e^{\eta_l^{(b+1)}}, e^{\phi_l^{(b+1)}})]^{1-\delta_{is}} [\int_{t_{is}}^\infty\Gamma(u_{is}\mid e^{\eta_l^{(b+1)}}, e^{\phi_l^{(b+1)}})du_{is}]^{\delta_{is}}\text{N}(x_{is}\mid \beta^{(b+1)}_l, \kappa^{2(b+1)}_l)/\{\sum_{l=1}^Lp_{ls}[ $ $\Gamma(t_{is}\mid e^{\eta_l^{(b+1)}}, e^{\phi_l^{(b+1)}})]^{1-\delta_{is}}[\int_{t_{is}}^\infty\Gamma(u_{is}\mid e^{\eta_l^{(b+1)}}, e^{\phi_l^{(b+1)}})du_{is}]^{\delta_{is}}\text{N}(x_{is}\mid  \beta^{(b+1)}_l, \kappa^{2(b+1)}_l)\}$ with $p_{1s} = 1-\zeta_{1s}$ and $p_{ls}=(1-\zeta_{ls})\prod_{r=1}^{l-1}\zeta_{rs}$ for $l = 2,..., L-1$.\\

For the update of ${\boldsymbol \mu} $ we have $p({\boldsymbol \mu}  \mid {\Psi}, data) \propto \text{N}_2({\boldsymbol \mu}\mid a_\mu, B_\mu)\prod_{l=1}^L \text{N}_2((\eta_l, \phi_l)'\mid {\boldsymbol \mu}, {\boldsymbol \Sigma})$, so we sample $p({\boldsymbol \mu}^{(b)}  \mid data, \Psi) \stackrel{\mbox{}}{\sim}  \text{N}_2(m_\mu, S_\mu^2)$ where $m_\mu = S_\mu^2(B_\mu^{-1}a_\mu + {\boldsymbol \Sigma}^{-1}\sum_{l=1}^L(\eta_l, \phi_l)'^{(b)})$, $ S_\mu^2 = (B_\mu^{-1} + L{\boldsymbol \Sigma}^{-1(b)})^{-1}$.  \\

For the update of ${\boldsymbol \Sigma}$, we have $p({\boldsymbol \Sigma} \mid {\Psi}, data) \propto \prod_{l=1}^L \text{N}_2((\eta_l, \phi_l)'\mid {\boldsymbol \mu}, {\boldsymbol \Sigma}) \text{IWish}({\boldsymbol \Sigma} \mid a_\Sigma, B_\Sigma)$, so we sample $p({\boldsymbol \Sigma}^{(b+1)} \mid data, \Psi) \stackrel{\mbox{}}{\sim} \text{IWish}(L + a_\Sigma, B_\Sigma + \sum_{l=1}^L ((\eta_l, \phi_l)'^{(b+1)} - {\boldsymbol \mu}^{(b+1)})((\eta_l, \phi_l)'^{(b+1)} - {\boldsymbol \mu}^{(b+1)})')$\\

For the update of $\lambda$ we have $p(\lambda \mid {\Psi}, data) \propto  \text{N}(\lambda \mid a_\lambda, b^2_\lambda)\prod_{l=1}^L \text{N}(\beta_l \mid \lambda, \tau^2)$, so we sample $p(\lambda^{(b+1)}\mid {\Psi}, data)  \stackrel{\mbox{}}{\sim} \text{N}(m_\lambda, s^2_\lambda)$ where $m_\lambda = s^2_\lambda(b_\lambda^{-2}a_\lambda + \tau^{-2(b)}\sum_{l=1}^L \beta_l^{(b+1)})$ and $s^2_\lambda = ( b^{-2}_\lambda + \tau^{-2(b)}L)^{-1}$. \\

For the update of $\tau^2$ we have $p(\tau^2 \mid {\Psi}, data) \propto  \Gamma^{-1}(\tau^2\mid a_\tau, b_\tau)\prod_{l=1}^L N(\beta_l \mid \lambda, \tau^2)$, so we sample $p(\tau^{2(b+1)} \mid {\Psi}, data)  \stackrel{\mbox{}}{\sim} \Gamma^{-1}(0.5L + a_\tau, 0.5[\sum_{l=1}^L (\beta_l^{(b+1)} - \lambda^{(b+1)})^2 ] + b^\tau)$ \\

For the update of $\rho$, $p(\rho \mid {\Psi}, data) \propto \Gamma(\rho\mid  a_\rho,b_\rho)\prod_{l=1}^L  \Gamma^{-1}(\kappa^2_l\mid a, \rho)$, so we sample $p(\rho^{(b+1)} \mid {\Psi}, data) \stackrel{\mbox{}}{\sim} \Gamma(aL+ a_\rho, [\sum_{l=1}^L  \kappa_l^{-2(b+1)}] + b_\rho)$. \\

We do not have conjugacy for $\alpha$ and $b$, so we turn to the Metropolis-Hastings algorithm to update these parameters.  The Bivariate Beta density of $(\zeta_{c}, \zeta_{T})$, has a complicated form, however, we can work with the density of the latent variables, $(U,V,W)$: $p(\alpha, b \mid {\Psi}, data) \propto  \text{Unif}(b\mid 0,1)\Gamma(\alpha\mid  a_\alpha, b_\alpha)\prod_{l-1}^{L-1}\text{Beta}(U_l\mid \alpha,1-b)\text{Beta}(V_l\mid \alpha, 1-b) \text{Beta}(W_l\mid 1+\alpha - b, b)$. We sample from the proposal distribution,  $(\log(\alpha'), \text{logit}(b'))' \sim \text{N}_2((\log(\alpha^{(b)}),\text{logit}(b^{(b)})), cS^2_{\alpha b})$, where $S_{\alpha b}^2 $ is updated from the average variances and covariance of posterior samples of $((\log(\alpha),\text{logit}(b))$ under initial runs, and $c $ is updated from initial runs to optimize mixing.

\vspace{6pt}
\subsection*{B.3 Conditional Predictive Ordinate Derivations} 
%\noindent {\Large \textbf{B.3 Conditional Predictive Ordinate Derivations}} 

\noindent Here we provide the details of how we arrived to the expression necessary for computing the CPO values under the DDP mixture model.  As our data example in Section 4.1 does not contain any random covariates, we will derive the expression without covariates, however, the derivation can easily be extended to include random covariates in the curve-fitting setting.  The hierarchical form of the  DDP mixture model without covariates and based on the truncation approximation, $G_{Ls}$, of $G_s$  is given as follows:
\begin{eqnarray*}
t_{is}|{\mathrm w}_{is}, {\boldsymbol \theta} &\stackrel{\text{ind}}{\sim}& \Gamma(t_{is}\mid   {\boldsymbol \theta}_{{\mathrm w}_{is}} )\  \ \mbox{for} \ i= 1,...,n_s \ \  s\in\{C,T\} \\
 {\boldsymbol {\mathrm w}}\mid \{\zeta_{lC}, \zeta_{lT}\} &\sim& \prod_{s\in \{C,T\}}\prod_{i=1}^{n_s}\sum_{l=1}^L\left[(1-\zeta_{ls})\prod_{r=1}^{l-1}\zeta_{rs}\right]\delta_l ({\mathrm w}_{is}) \\
 {\boldsymbol \theta}_l \mid {\boldsymbol \mu}, {\boldsymbol \Sigma} &\stackrel{\text{i.i.d.}}{\sim}& \text{N}_2( {\boldsymbol \theta}_l\mid {\boldsymbol \mu}, {\boldsymbol \Sigma}) \\
 (\zeta_{lC},\zeta_{lT})\mid \alpha , b &\stackrel{\text{i.i.d.}}{\sim}& \text{Biv-Beta}( (\zeta_{lC},\zeta_{lT})\mid \alpha , b) \  \ \mbox{for} l= 1,...,L-1 
 \end{eqnarray*}
with $\alpha \sim \Gamma(\alpha\mid a_\alpha,b_\alpha)$, $b\sim \text{Unif}(b\mid 0,1)$, ${\boldsymbol \mu}\sim \text{N}_2({\boldsymbol \mu}\mid a_\mu, B_\mu)$, and ${\boldsymbol \Sigma}\sim \text{IWish}({\boldsymbol \Sigma}\mid a_\Sigma, B_\Sigma)$.  Let $\Psi = (\alpha, b, {\boldsymbol \mu}, {\boldsymbol \Sigma})$. The predictive density for a new survival time from group $s$, $t_{0s}$, is given by:
\begin{eqnarray*}
p(t_{0s}\mid data) &=& \int\int\Gamma(t_{0s}\mid  {\boldsymbol \theta}_{{\mathrm w}_{0s}})\left(\sum_{l=1}^Lp_{ls}\delta_l({\mathrm w}_{0s})\right)p({\boldsymbol \theta},{\boldsymbol p}, {\boldsymbol {\mathrm w}}, \Psi \mid data)d{\mathrm w}_{0s}d{\boldsymbol \theta}d{\boldsymbol {\mathrm w}} d{\boldsymbol p} d\Psi \\
 &=& \int \left(\sum_{l=1}^L p_{ls}\Gamma(t_{0s}\mid {\boldsymbol \theta}_l)\right)p({\boldsymbol \theta},{\boldsymbol p}, {\boldsymbol {\mathrm w}}, \Psi \mid data)d{\boldsymbol \theta}d{\boldsymbol {\mathrm w}} d{\boldsymbol p} d\Psi
\end{eqnarray*}
\indent Let $s'$ be the experimental group that $s$ is not, $data = \{{\boldsymbol t}_s, {\boldsymbol t}_{s'}\}$, and $A$ be the normalizing constant for $p({\boldsymbol \theta},{\boldsymbol p}, {\boldsymbol {\mathrm w}}, \Psi \mid data)$.  Namely, $p({\boldsymbol \theta},{\boldsymbol p}, {\boldsymbol {\mathrm w}}, \Psi \mid data) = [(\prod_{i=1}^{n_s} \Gamma(t_{is}\mid {\boldsymbol \theta}_{{\mathrm w}_{is}})) (\prod_{i=1}^{n_{s'}} \Gamma(t_{is'}\mid {\boldsymbol \theta}_{{\mathrm w}_{is'}}))p({\boldsymbol \theta},{\boldsymbol p}, {\boldsymbol {\mathrm w}}, \Psi)] / [\int (\prod_{i=1}^{n_s} \Gamma(t_{is}\mid {\boldsymbol \theta}_{{\mathrm w}_{is}}))(\prod_{i=1}^{n_{s'}} \Gamma(t_{is'}\mid  {\boldsymbol \theta}_{{\mathrm w}_{is'}}))p({\boldsymbol \theta},{\boldsymbol p}, {\boldsymbol {\mathrm w}}, \Psi)$ $ d{\boldsymbol \theta}d{\boldsymbol {\mathrm w}} d{\boldsymbol p} d\Psi]$. Note, $p({\boldsymbol \theta},{\boldsymbol p}, {\boldsymbol {\mathrm w}}, \Psi) = \text{N}_2({\boldsymbol \theta}\mid  {\boldsymbol \mu}, {\boldsymbol \Sigma})(\prod_{i=1}^{n_{s}}\sum_{l=1}^L p_{ls} \delta_l({\mathrm w}_{is})) (\prod_{i=1}^{n_{s'}}\sum_{l=1}^L p_{ls'} $ $\delta_l({\mathrm w}_{is'}))$ $  \text{Biv-Beta}({\boldsymbol p}\equiv (\zeta_s,\zeta_{s'})\mid  \alpha, b) \Gamma(\alpha\mid a_\alpha, b_\alpha)\text{Unif}(b\mid 0,1)N_2({\boldsymbol \mu}\mid a_\mu,B_\mu) \text{IWish}({\boldsymbol \Sigma}\mid a_\Sigma, $ $B_\Sigma)$. 

The CPO of the $ith$ survival time in group $s$ is defined as, $\text{CPO}_{is} = p(t_{is}\mid {\boldsymbol t}_{(-i)s}, {\boldsymbol t}_{s'}) = \int \Gamma(t_{is}\mid {\boldsymbol \theta}_{{\mathrm w}_{0s}})(\sum_{l=1}^L p_{ls}\delta_l({\mathrm w}_{0s}))p({\boldsymbol \theta},{\boldsymbol p}, {\boldsymbol {\mathrm w}}_{(-i)s}, \Psi) d{\boldsymbol \theta}d{\boldsymbol {\mathrm w}_{(-i)s}} d{\boldsymbol p} d\Psi d{\mathrm w}_{0s}$, where ${\boldsymbol {\mathrm w}}_{(-i)s}$ is the vector ${\boldsymbol {\mathrm w}}$ with the $i^{th}$ member of group $s$ removed. Similarly, $data_{(-i)s}$ represents $data$ with the $i^{th}$ member in group $s$ removed.  Now, consider $p({\boldsymbol \theta}, {\boldsymbol p}, {\boldsymbol {\mathrm w}}_{(-i)s}, \Psi|data_{(-i)s})$, which is given by:
\begin{eqnarray*}
&& \frac{p(data_{(-i)s}\mid {\boldsymbol \theta},{\boldsymbol {\mathrm w}}_{(-i)s}) p({\boldsymbol \theta}, {\boldsymbol {\mathrm w}}_{(-i)s}, {\boldsymbol p},\Psi)}{\int p(data_{(-i)s}\mid {\boldsymbol \theta}, {\boldsymbol {\mathrm w}}_{(-i)s}) p({\boldsymbol \theta}, {\boldsymbol {\mathrm w}}_{(-i)s}, {\boldsymbol p},\Psi)d{\boldsymbol {\mathrm w}_{(-i)s}}  d{\boldsymbol p} d\Psi} \\
&&= \frac{\left\{\prod_{j\neq i}^{n_s} \Gamma(t_{js}\mid {\boldsymbol \theta}_{{\mathrm w}_{js}})\right\}\left\{\prod_{i=1}^{n_{s'}} \Gamma(t_{is'}\mid {\boldsymbol \theta}_{{\mathrm w}_{is'}})\right\}p({\boldsymbol \theta},{\boldsymbol p}, {\boldsymbol {\mathrm w}}_{(-i)s}, \Psi)}{\int \left\{\prod_{j\neq i}^{n_s} \Gamma(t_{js}\mid {\boldsymbol \theta}_{{\mathrm w}_{js}})\right\}\left\{\prod_{i=1}^{n_{s'}} \Gamma(t_{is'}\mid {\boldsymbol \theta}_{{\mathrm w}_{is'}})\right\}p({\boldsymbol \theta},{\boldsymbol p}, {\boldsymbol {\mathrm w}}_{(-i)s}, \Psi) d{\boldsymbol \theta}d{\boldsymbol {\mathrm w}}_{(-i)s} d{\boldsymbol p} d\Psi}
\end{eqnarray*}
Let $B_{is}$ be the normalizing constant of $p({\boldsymbol \theta}, {\boldsymbol p}, {\boldsymbol {\mathrm w}}_{(-i)s}, \Psi\mid data_{(-i)s})$, specifically:\\ 
$B_{is} = \int \left\{\prod_{j\neq i}^{n_s} \Gamma(t_{js}\mid {\boldsymbol \theta}_{{\mathrm w}_{js}})\right\}\left\{\prod_{i=1}^{n_{s'}} \Gamma(t_{is'}\mid {\boldsymbol \theta}_{{\mathrm w}_{is'}})\right\}p({\boldsymbol \theta},{\boldsymbol p}, {\boldsymbol {\mathrm w}}_{(-i)s}, \Psi) d{\boldsymbol \theta}d{\boldsymbol {\mathrm w}}_{(-i)s} d{\boldsymbol p} d\Psi$\\
Then, we can write $p({\boldsymbol \theta}, {\boldsymbol p}, {\boldsymbol {\mathrm w}}_{(-i)s}, \Psi|data_{(-i)s}) $ as:
\begin{eqnarray*}
 &&\frac{\left\{\prod_{i=1}^{n_s} \Gamma(t_{is}\mid {\boldsymbol \theta}_{{\mathrm w}_{is}})\right\}\left\{\prod_{i=1}^{n_{s'}} \Gamma(t_{is'}\mid {\boldsymbol \theta}_{{\mathrm w}_{is'}})\right\}p({\boldsymbol \theta},{\boldsymbol p}, {\boldsymbol {\mathrm w}}, \Psi)}{B_{is} \Gamma(t_{is}\mid {\boldsymbol \theta}_{{\mathrm w}_{is}}) p({\mathrm w}_{is}\mid {\boldsymbol p})}=\frac{A}{B_{is}} \frac{ p({\boldsymbol \theta},{\boldsymbol p}, {\boldsymbol {\mathrm w}}, \Psi\mid data)}{\Gamma(t_{is} \mid {\boldsymbol \theta}_{{\mathrm w}_{is}}) p({\mathrm w}_{is}\mid{\boldsymbol p})}
\end{eqnarray*}
Thus, 
\begin{eqnarray*}
\text{CPO}_{is} &=& \int \Gamma(t_{is}\mid {\boldsymbol \theta}_{{\mathrm w}_{0s}})p({\mathrm w}_{0s}\mid {\boldsymbol p})p({\boldsymbol \theta},{\boldsymbol p}, {\boldsymbol {\mathrm w}}_{(-i)s}, \Psi) d{\boldsymbol \theta}d{\boldsymbol {\mathrm w}_{(-i)s}} d{\boldsymbol p} d\Psi d{\mathrm w}_{0s} \\
&=& \int \Gamma(t_{is}\mid {\boldsymbol \theta}_{{\mathrm w}_{0s}}) \left( \int p( {\mathrm w}_{0s},{\mathrm w}_{is}\mid {\boldsymbol p})d{\mathrm w}_{is} \right)p({\boldsymbol \theta},{\boldsymbol p}, {\boldsymbol {\mathrm w}}_{(-i)s}, \Psi) d{\boldsymbol \theta}d{\boldsymbol {\mathrm w}_{(-i)s}} d{\boldsymbol p} d\Psi d{\mathrm w}_{0s}\\
&=& \frac{A}{B_{is}}\int \frac{\Gamma(t_{is} \mid {\boldsymbol \theta}_{{\mathrm w}_{0s}}) p({\mathrm w}_{0s},{\mathrm w}_{is}\mid {\boldsymbol p})}{\Gamma(t_{is} \mid {\boldsymbol \theta}_{{\mathrm w}_{is}}) p({\mathrm w}_{is}\mid {\boldsymbol p})}p({\boldsymbol \theta}, {\boldsymbol p}, {\boldsymbol {\mathrm w}}, \Psi\mid data) d{\mathrm w}_{0s}d{\boldsymbol \theta }d{\boldsymbol {\mathrm w} d{\boldsymbol p} d\Psi }\\
&=& \frac{A}{B_{is}} \int \frac{\sum_{l=1}^L p_{ls} \Gamma(t_{is} \mid {\boldsymbol \theta}_l)}{\Gamma(t_{is}\mid  {\boldsymbol \theta}_{{\mathrm w}_{is}})} p({\boldsymbol \theta}, {\boldsymbol p}, {\boldsymbol {\mathrm w}}, \Psi\mid data) d{\mathrm w}_{0s}d{\boldsymbol \theta}d{\boldsymbol {\mathrm w}} d{\boldsymbol p} d\Psi
\end{eqnarray*}

Note, $p({\mathrm w}_{0s}\mid {\mathrm w}_{is},{\boldsymbol p}) = p({\mathrm w}_{0s}\mid {\boldsymbol p}) $. All that is left is to be able to evaluate $A/B_{is}$:
\begin{eqnarray*}
\left(\frac{A}{B_{is}}\right)^{-1} &=& \frac{1}{A} \int \left\{\prod_{j\neq i}^{n_s}\Gamma(t_{js}\mid {\boldsymbol \theta}_{{\mathrm w}_{js}})\right\} \left\{\prod_{i=1}^{n_{s'}}\Gamma(t_{is'}\mid  {\boldsymbol \theta}_{{\mathrm w}_{is'}})\right\}\underbrace{\left(\int p({\mathrm w}_{is}\mid{\boldsymbol {\mathrm w}}_{(-i)s}, {\boldsymbol p})d{\mathrm w}_{is}\right)}_{1} \\
&& \times p({\boldsymbol {\mathrm w}}_{(-i)s} \mid {\boldsymbol p}) p({\boldsymbol p}, {\boldsymbol \theta},\Psi) d{\boldsymbol \theta}d{\boldsymbol {\mathrm w}}_{(-i)s} d{\boldsymbol p} d\Psi  \\
&=&  \frac{1}{A} \int \left\{\prod_{j\neq i}^{n_s}\Gamma(t_{js}\mid  {\boldsymbol \theta}_{{\mathrm w}_{js}})\right\} \left\{\prod_{i=1}^{n_{s'}}\Gamma(t_{is'}\mid {\boldsymbol \theta}_{{\mathrm w}_{is'}})\right\}p({\boldsymbol \theta}, {\boldsymbol p}, {\boldsymbol {\mathrm w}}, \Psi) d{\boldsymbol \theta}d{\boldsymbol {\mathrm w}} d{\boldsymbol p} d\Psi \\
&=&  \frac{1}{A} \int \frac{\left\{\prod_{j\neq i}^{n_s}\Gamma(t_{js}\mid {\boldsymbol \theta}_{{\mathrm w}_{js}})\right\} \left\{\prod_{i=1}^{n_{s'}}\Gamma(t_{is'}\mid  {\boldsymbol \theta}_{{\mathrm w}_{is'}})\right\}}{\Gamma(t_{is}\mid{\boldsymbol \theta}_{{\mathrm w}_{is}})}p({\boldsymbol \theta}, {\boldsymbol p}, {\boldsymbol {\mathrm w}}, \Psi) d{\boldsymbol \theta}d{\boldsymbol {\mathrm w}} d{\boldsymbol p} d\Psi\\
&=& \int \frac{1}{\Gamma(t_{is}\mid  {\boldsymbol \theta}_{{\mathrm w}_{is}})}p({\boldsymbol \theta}, {\boldsymbol p}, {\boldsymbol {\mathrm w}}, \Psi) d{\boldsymbol \theta}d{\boldsymbol {\mathrm w}} d{\boldsymbol p} d\Psi
\end{eqnarray*}
In summary,
\begin{eqnarray*}
\text{CPO}_{is} &=& \frac{A}{B_{is}}\int\frac{  \sum_{l=1}^L p_{ls} \Gamma(t_{is} \mid {\boldsymbol \theta}_l)}{\Gamma(t_{is}\mid  {\boldsymbol \theta}_{{\mathrm w}_{is}})} p({\boldsymbol \theta}, {\boldsymbol p}, {\boldsymbol {\mathrm w}}, \Psi| data) d{\mathrm w}_{0s}d{\boldsymbol \theta}d{\boldsymbol {\mathrm w}} d{\boldsymbol p} d\Psi \\
&& \text{where} \ \left(\frac{A}{B_{is}}\right)^{-1} =  \int \frac{1}{\Gamma(t_{is}\mid {\boldsymbol \theta}_{{\mathrm w}_{is}})}p({\boldsymbol \theta}, {\boldsymbol p}, {\boldsymbol {\mathrm w}}, \Psi) d{\boldsymbol \theta}d{\boldsymbol {\mathrm w}} d{\boldsymbol p} d\Psi
\end{eqnarray*}
The MCMC approximation of the CPO values is given by:
\begin{eqnarray*}
\text{CPO}_{is} &\approx& \frac{A}{B_{is}}\left( \sum_{j=1}^B \frac{\sum_{l=1}^L p_{ls}\Gamma(t_{is} \mid {\boldsymbol \theta}_{lj})}{\Gamma(t_{is} \mid {\boldsymbol \theta}_{{\boldsymbol {\mathrm w}}_{isj}})}\right),  \mbox{ \ \ where \ \ }  \frac{A}{B_{is}} = \left( \sum_{j=1}^B \frac{1}{\Gamma(t_{is} \mid {\boldsymbol \theta}_{{\boldsymbol {\mathrm w}}_{isj}})}\right) 
\end{eqnarray*}
where $B$ is the total number of MCMC iterations.

\bibliographystyle{bka}

\bibliography{biomsample}

\end{document}